\documentclass[prd,notitlepage,superscriptaddress]{revtex4-1}

\usepackage{physics}
\usepackage{amsmath}
\usepackage{amssymb}  
\usepackage{graphicx}  
\usepackage{verbatim} 
\usepackage{color}    
\usepackage{subfigure}  
\usepackage{hyperref}  
\usepackage{mathrsfs}
\usepackage{float}

\begin{document}

\title{Spin-Independent Two-Neutrino Exchange Potential with Mixing and \boldmath $CP$-Violation}
\date{\today}
\author{Quan Le Thien}
\affiliation{Physics Department, Wabash College, Crawfordsville, IN 47933, USA}
\author{Dennis E. Krause}
\email{kraused@wabash.edu}
\affiliation{Physics Department, Wabash College, Crawfordsville, IN 47933, USA}
\affiliation{Department of Physics  and Astronomy, Purdue University, West Lafayette, IN 47907, USA}

\begin{abstract}
We develop a new approach for calculating the spin-independent two-neutrino exchange potential (2NEP) between nonrelativistic fermions which places emphasis on the neutrino vacuum state, an area of theoretical interest in recent years.  The 2NEP is a natural probe of fundamental issues of neutrino physics such as neutrino masses, flavor mixing, the number of neutrino flavors, neutrino nature (Dirac or Majorana), $CP$-violation, and the neutrino vacuum state. We explore the dependence of the 2NEP on the mixing of neutrino mass states assuming normal and inverted mass ordering for nucleon-nucleon, nucleon-lepton, and lepton-lepton interactions, and  the $CP$-violation phase in the neutrino mixing matrix.

\end{abstract}

\maketitle

\section{Introduction}

Of all the parts of the Standard Model, the neutrino sector arguably holds the greatest promise for revealing new physics beyond the Standard Model.  Since the discovery of the electron-neutrino over sixty years ago \cite{Cowan}, neutrinos have proven to be a constant source of surprise, from the discovery of three different flavors to the phenomenon of neutrino oscillations \cite{Bilenky}.  Yet, many basic properties of neutrinos remain unknown, including their masses and whether neutrinos and antineutrinos are distinct particles (i.e., whether they are Dirac or Majorana fermions).  The recent paper by Stadnik \cite{Stadnik} has drawn renewed attention to a lesser known aspect of neutrinos:  the virtual exchange of neutrino-antineutrino pairs leads to weak long-range forces.   While the magnitudes of these forces are quite small, making their observation difficult, they remain of theoretical interest because of their unique nature: the virtual exchange requires that {\em all} neutrino properties and energies must contribute in some way to these forces.  The purpose of this paper is to open a new avenue of exploration of neutrino properties using neutrino exchange forces by incorporating mixing of mass states which includes $CP$-violation.

The observation of neutrino oscillations \cite{2015 Nobel Prize} not only demonstrates that neutrinos have mass, but also that the three different mass states mix, a phenomenon previously observed in neutral mesons \cite{Barr}.   In addition, the possibilities of neutron-antineutron oscillations \cite{Phillips} and oscillations of neutral atoms  \cite{Bernabeu} have been investigated in theories which violate baryon number and total lepton number conservation, respectively.  The quantum field theory description of particle mixing has been an area of intensive study \cite{Beuthe,Capolupo}. A number of different approaches have been applied to neutrino mixing \cite{Shrock PLB,Shrock PRD,Beuthe,Capolupo,Akhmedov,Kruppke,Giunti,Ho,Nishi 2006,Nishi 2008,Bernardini}. The most straightforward method \cite{Shrock PLB,Shrock PRD}  only assumes the neutrino flavor fields appearing in the weak interaction vertices are linear combinations of neutrino mass fields, while other field-theoretic approaches seek to define neutrino flavor fields explicitly in terms of the different flavors' creation and annihilation operators, ultimately suggesting new nontrivial structures of neutrino vacua \cite{BV AoP,BV PLB,BV 2019,Tureanu}. However, one finds that all of these approaches lead to similar results when applied to experiments involving ultrarelativistic neutrinos which are encountered in actual experiments \cite{BV PLB,Lee}.  (When cosmological arguments are combined with measurements from neutrino oscillation experiments, one finds that neutrino masses are $\lesssim 0.1$~eV, which is much smaller than observed neutrino energies.) To find differences between various theories one needs to find phenomena involving coherent nonrelativistic neutrinos. In this paper, we will discuss a situation which clearly requires a quantum field theory approach for neutrinos and which must include all possible neutrino flavors and energies: the two-neutrino exchange potential (2NEP) with mixing.

 To accomplish this, we will first quickly review previous derivations of the 2NEP.  Then we will introduce an alternative method for calculating the single flavor 2NEP based on expressing the neutrino fields in the Schr\"{o}dinger picture and using  time-independent perturbation theory to evaluate the shift in neutrino field vacuum energy due to the presence of two fermions.   This approach has been used previously to calculate the Yukawa interaction between fermions exchanging scalar bosons \cite{Wentzel,March,Strocchi,MR}, the spin-dependent interaction due to pseudoscalar exchange \cite{March,GT}, and the electrostatic Coulomb interaction between two charged particles arising from photon exchange \cite{CT}. We take this more noncovariant approach  because  it places an explicit emphasis on the neutrino vacuum, and  it separates out the effects of neutrino mixing in the weak interaction Hamiltonian from those due to the nontrivial neutrino vacua. Hence, we believe that this method allows our results to arise more transparently than other modern approaches for calculating potentials.  After presenting our approach, we proceed to generalize the single flavor result by including the charged current interaction and incorporating three neutrino flavors through the mixing of neutrino fields using the Pontecorvo-Maki-Nakagawa-Sakata (PMNS) matrix, and investigate the resulting potential for various combinations of fermions.  We consider how the potentials depend on: (1) the use of  normal and inverted ordering of neutrino masses, including the case when one of the neutrino mass states is massless, and (2) $CP$-violation in the PMNS matrix.   We conclude the paper with a summary of our results and a discussion of how this work can be extended.

\section{Previous 2NEP Derivations}

\begin{figure}[b]
\begin{center}
\includegraphics[height=1.5in]{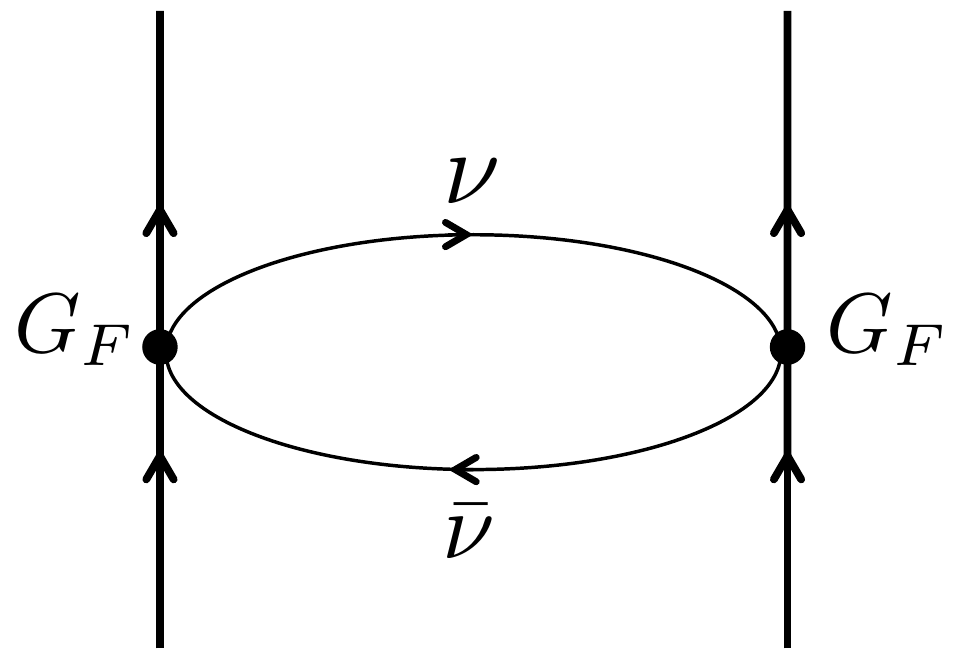}
\caption{Feynman diagram of neutrino-antineutrino exchange between fermions which  leads to Eq.~(\ref{FS V}).  The thicker (thinner) lines denote fermion (neutrino) propagators.}
\label{2NEP diagram}
\end{center}
\end{figure}

The possibility that a long-range force could arise from the exchange of virtual neutrinos has been known since the 1930s \cite{FS}, but it may have gained more widespread attention through Feynman's attempt to use virtual neutrino exchange to explain the force of gravity \cite{Feynman gravity}.   The first modern derivation of the 2NEP was done by Feinberg and Sucher \cite{FS} who applied dispersion-theoretic techniques to the low-energy four-fermion weak interaction.  For two electrons  separated by a distance $r$ exchanging a massless virtual neutrino-antineutrino pair (Fig.~\ref{2NEP diagram}), they found
\begin{equation}
V_{\nu,\bar{\nu}}(r) = \frac{G_{F}^{2}}{4\pi^{3}r^{5}},
\label{FS V}
\end{equation}
where $G_{F}$ is the Fermi constant.  (Throughout this paper, we assume units where $\hbar = c = 1$.)  They later recalculated the result using the Standard Model neutral current interaction, obtaining \cite{FSA},
\begin{equation}
V_{\nu,\bar{\nu}}(r) = \left(2\sin^{2}\theta_{W} + \frac{1}{2}\right)^{2}\frac{G_{F}^{2}}{4\pi^{3}r^{5}},
\label{FS SM}
\end{equation}
where $\theta_{W}$ is the Weinberg angle.  Hsu and Sikivie \cite{Hsu} obtained Eq.~(\ref{FS V}) using a more standard approach based on Feynman diagrammatic methods and Fourier transforms.  More recently, Segerra  has also written on the 2NEP using the dispersion theory approach \cite{Segarra}.

Shortly after Feinberg and Sucher's publication \cite{FS}, Hartle~\cite{Hartle} investigated the effects of multibody neutrino exchange forces in cosmology using a new  technique based on a formula obtained by Schwinger \cite{Schwinger}.  Later Fischbach, et al. used the Hartle-Schwinger formalism to derive Eq.~(\ref{FS SM}) and calculate its contribution to nuclear binding energies for tests of violations of the weak equivalence principle for the weak interaction \cite{Fischbach PRD}.    Fischbach then applied the multibody formalism to neutron stars, arguing that neutrino interactions lead to a catastrophic contribution to the star's self-energy unless neutrinos have a finite mass $\gtrsim 0.4$~eV \cite{Fischbach AoP}.  In the process of this work, he extended Eq.~(\ref{FS V}) to the case where a neutrino has mass $m_{\nu}$.   Shortly thereafter, Grifols, et al. \cite{Grifols} applied dispersion methods to the massive neutrino calculation to obtain (after adjusting notation) Fischbach's result in a more compact form,
\begin{equation}
V_{\nu,\bar{\nu}}(r) = \frac{G_{F}^{2}m_{\nu}^{3}}{4\pi^{3}r^{2}}K_{3}(2m_{\nu}r).
\label{massive V}
\end{equation}
They also calculated the 2NEP for Majorana neutrinos; all previous work assumed Dirac neutrinos.

Before continuing, a comment should be made on referring to the 2NEP as being ``long-ranged.''  For light neutrinos, $V_{\nu,\bar{\nu}}(r) \sim (1/r)(1/M_{Z}r)^{4}$, where $M_{Z}$ is the mass of the $Z^{0}$ boson.  Clearly $V_{\nu,\bar{\nu}}(r)$ falls off rapidly with separation due to the factor $(M_{Z}r)^{-4}$, but this  falloff is less rapid than one would expect for a weak interaction potential arising from the exchange of $Z_{0}$, which has a Yukawa potential form 
with range $M_{Z}^{-1} \sim 10^{-18}$~m.  Also, as emphasized recently by Asaka, et al. \cite{Asaka}, the 2NEP is based upon the low-energy Fermi approximation of the electroweak interaction and becomes invalid at short ranges where additional Standard Model contributions not accounted for in the low-energy approximation become important.  One expects deviations from Eq.~(\ref{massive V}) when $r \sim M_{Z}^{-1}$.  Furthermore, one should remember that a potential description of an interaction is a nonrelativistic concept and becomes of limited validity when the particle separation is less than their Compton wavelengths.

All of the above derivations of the 2NEP assume a single flavor of neutrinos, but mixing of neutrino flavors has now been observed experimentally \cite{2015 Nobel Prize} and its effects should be included.   Lusignoli and Petrarca \cite{LP} obtained an integral expression for the 2NEP including neutrino mixing using the Feinberg-Sucher method, but  did not obtain a  closed form result or include the interference of neutral current and charged current weak interactions.  Here we obtain a more detailed solution of the spin-independent 2NEP between fermions with flavor mixing of Dirac neutrinos and explore some of the consequences.

\section{Derivation of the Single Flavor two-neutrino Exchange Potential}

\subsection{Overview}

In order to establish our notation and approach for the mixing case, in this section  we derive the 2NEP by calculating the change in the single neutrino field vacuum energy due to the presence of two fermions which couple to the field.  The potential energy of these two particles is defined as the shift in the vacuum energy that depends on the separation of the particles.  In general, the field interaction will also lead to self-energy corrections which are independent of the particle separation.  

In our calculation, we will assume the two fermions  are located at positions $\vec{r}_{1}$ and $\vec{r}_{2}$ and are moving with negligible velocities and interacting with a neutrino field $\nu(\vec{r})$.  Since the system is nonrelativistic and a static potential is independent of time, we  adopt the noncovariant Schr\"{o}dinger picture for the fields, which is natural for this problem. The neutrino fields then have no time dependence which permits the use of ordinary time-independent Rayleigh-Schr\"{o}dinger perturbation theory to derive the potential.     We have not adopted more modern approaches based on calculating scattering amplitudes using Feynman diagrams (e.g., Refs.~\cite{Sakurai,Peskin}) for several reasons.  First, we wish to explicitly examine the dependence of the potential on the vacuum state since alternative vacuum states have been proposed for mixed neutrinos.  Second, since one is working directly in position space rather than momentum space, it is easier to isolate finite quantities, that depend on the separation distance between fermions which contribute to the interaction potential, from infinite self-energy terms, which do not.  Finally, it is straightforward to generalize our derivation of the single neutrino flavor 2NEP  to the case of three neutrinos with mixing.

\subsection{Hamiltonian}
\label{single neutrino H section}

We begin by writing the total Hamiltonian describing the two fermions interacting with the neutrino field as 
\begin{equation}
H = H_{0} + H_{\rm int},
\end{equation}
where $H_{0}$ is the free Hamiltonian describing the free neutrino field, and $H_{\rm int}$ describes the interaction between the fermions and neutrino field.  For the derivation of the static potential, the fermions have no dynamical properties themselves so the free Hamiltonian involves only the neutrino field,
\begin{equation}
H_{0} = \sum_{\vec{k},s}\omega_{\vec{k}}\left(b^{\dag}_{\vec{k},s}b_{\vec{k},s} + d^{\dag}_{\vec{k},s}d_{\vec{k},s}\right),
\label{single neutrino H0}
\end{equation}
where $b_{\vec{k},s}^{\dag}$ and $b_{\vec{k},s}$ are the creation and annihilation operators for neutrinos with momentum $\vec{k}$ and spin state $s$, and $d_{\vec{k},s}^{\dag}$ and $d_{\vec{k},s}$ are the corresponding antineutrino creation and annihilation operators (i.e., we are assuming the neutrinos are Dirac neutrinos).  For neutrinos of mass $m_{\nu}$, $\omega_{\vec{k}}^{2} = m_{\nu}^{2} + \vec{k}^{2}$; for a single neutrino, there is no distinction between the mass and flavor fields.
In Eq.~(\ref{single neutrino H0}), the infinite free field vacuum energy has been subtracted away, so the free field vacuum energy to be used in subsequent calculations vanishes: $E^{(0)}_{\rm vac} = 0$.  

To facilitate a comparison of our results with previous derivations of the 2NEP, in this section we will assume the interaction Hamiltonian $H_{\rm int}$ is  due entirely to the neutral current weak interaction, which, for a single neutrino flavor, is given by
\begin{equation}
H_{\rm int} = H_{\rm int}^{\rm NC} = \frac{G_{F}}{\sqrt{2}} \int d^{3}r \,J^{f}_{\mu}(\vec{r})\,\left[\bar{\nu}(\vec{r})\gamma^{\mu}\left(1 - \gamma^{5}\right)\nu(\vec{r})\right].
\label{H int general}
\end{equation}
\begin{table}[t]
\caption{Values  for the vector coupling $g_{V}^{f}$ for  neutral current (NC) lepton-neutrino and nucleon-neutrino interactions, where $\theta_{W}$ is the Weinberg angle and $\sin^{2}\theta_{W} = 0.2223$ \cite{PDG 2018}.}
\begin{center}
\begin{tabular}{ccc}\hline\hline
Fermion &&  $g_{V}^{f}$  \\ \hline
$e$, $\mu$, $\tau$ && $\frac{1}{2} + 2 \sin^{2}\theta_{W}$  \\
Proton &&  $\frac{1}{2} - 2 \sin^{2}\theta_{W}$ \\ 
Neutron && $-\frac{1}{2}$  \\ \hline\hline
\end{tabular}
\end{center}
\label{g table}
\end{table}%
Here $J^{f}_{\mu}(\vec{r})$ is the fermion current
\begin{equation}
J^{f}_{\mu}(\vec{r}) =  \bar{f}(\vec{r})\gamma_{\mu}\left(g_{V}^{f} - g_{A}^{f}\gamma^{5}\right)f(\vec{r}),
\label{Jf}
\end{equation}
where $f(\vec{r})$ and $\bar{f}(\vec{r})$ are the fermion and antifermion fields, and $g_{V}^{f}$ and $g_{A}^{f}$ are the vector and axial couplings, respectively.  In Eq.~(\ref{Jf}), $\nu(\vec{r})$ and $\bar{\nu}(\vec{r})$ are the neutrino and antineutrino fields, which in the Schr\"{o}dinger picture are given by 
\begin{subequations}
\begin{eqnarray}
\nu(\vec{r}) & = &  \sum_{\vec{k},s}\left(\frac{m_{\nu}}{V\omega_{\vec{k}}}\right)^{1/2}
	\left[b_{\vec{k},s}u_{s}(\vec{k})e^{i\vec{k} \cdot \vec{r}} + d^{\dag}_{\vec{k},s}v_{s}(\vec{k})e^{-i\vec{k}\cdot \vec{r})}\right],
\label{psi} \\
\nu^{\dag}(\vec{r}) & = &  \sum_{\vec{k},s}\left(\frac{m_{\nu}}{V\omega_{\vec{k}}}\right)^{1/2}
	\left[b_{\vec{k},s}^{\dag}u^{\dag}_{s}(\vec{k})e^{-i\vec{k} \cdot \vec{r}} + d_{\vec{k},s}v^{\dag}_{s}(\vec{k})e^{i\vec{k}\cdot \vec{r}}\right],
\label{psi bar}
\end{eqnarray}
\end{subequations}
where $V$ is the normalization volume and $\nu^{\dag}(\vec{r}) = \bar{\nu}(\vec{r}) \gamma^{0}.$   For our two static fermions which are essentially classical particles, the spin-independent fermion current comes from the vector portion of the fermion current,
\begin{equation}
J_{\mu}^{f}(\vec{r}) = J_{0}^{f}(\vec{r})\delta_{\mu,0} =  \left[g_{V,1}^{f}\delta^{3}(\vec{r} - \vec{r}_{1}) +  g_{V,2}^{f}\delta^{3}(\vec{r} - \vec{r}_{2})\right]\delta_{\mu,0},
\label{rho}
\end{equation}
where $\vec{r}_{i}$ is the position of the $i$th fermion.  (The axial portion of the fermion current would lead to a spin-dependent 2NEP which is beyond the scope of this paper.)  The values for the vector couplings $g_{V}^{f}$ for leptons and nucleons are given in Table~\ref{g table}. 
For antifermions, the corresponding current would have opposite sign: $J_{\mu}^{\bar{f}}(\vec{r})  = -J_{\mu}^{f}(\vec{r}) $.
The total interaction Hamiltonian  with the two fermions  then becomes
\begin{equation}
H_{\rm int} = \sum_{i = 1}^{2}H_{{\rm int},i},
\label{2 particle H int}
\end{equation}
where $H_{{\rm int},i}$ is the interaction Hamiltonian involving the $i$th fermion, which is given by

\begin{equation}
H_{{\rm int},i} = \frac{G_{F} g_{V,i}^{f}}{\sqrt{2}}\nu^{\dag}(\vec{r}_{i}) \left(1 - \gamma^{5}\right) \nu(\vec{r}_{i}).
\label{H int i}
\end{equation}

\subsection{Perturbation Theory Calculation}

We will now use ordinary time-independent perturbation theory to calculate the energy of the system due to the interaction.  The first-order energy correction of the ground state does not lead to a potential energy between the particles because it is the sum of two terms, each of which only depends on the position of one particle:
\begin{equation}
E_{\rm vac}^{(1)} = \langle 0|H_{\rm int}|0\rangle = \langle 0|H_{\rm int,1}|0\rangle + \langle 0|H_{\rm int,2}|0\rangle
 = E_{\rm vac}^{(1)}(\vec{r}_{1}) + E_{\rm vac}^{(1)}(\vec{r}_{2}),
\end{equation}
where $|0\rangle$ is the unperturbed vacuum state of the neutrino field, and  $E^{(1)}_{\rm vac}(\vec{r}_{i})$ is the first-order self-energy of the $i$th particle.

The 2NEP must come from the second-order ground state energy correction which can be written as
\begin{equation}
E_{\rm vac}^{(2)} = \sum_{E^{(0)}_{n}\neq E^{(0)}_{\rm vac}}
	\frac{\langle 0|H_{\rm int}|E^{(0)}_{n}\rangle \langle E^{(0)}_{n}|H_{\rm int}|0\rangle}{E_{\rm vac}^{(0)} - E^{(0)}_{n}}
	= \sum_{E^{(0)}_{n}\neq 0}\frac{\langle 0|H_{\rm int}|E^{(0)}_{n}\rangle \langle E^{(0)}_{n}|H_{\rm int}|0\rangle}{-E^{(0)}_{n}},
\label{E vac 2a}
\end{equation}
where  $|E^{(0)}_{n}\rangle$ is a non-vacuum state with energy $E^{(0)}_{n}$, and we have set  the zero-point constant energy of the free field vacuum $E_{\rm vac}^{(0)} = 0$.  When Eq.~(\ref{2 particle H int}) is substituted into Eq.~(\ref{E vac 2a}), we find three types of terms:
\begin{equation}
E_{\rm vac}^{(2)} = E_{\rm vac}^{(2)}(\vec{r}_{1}) + E_{\rm vac}^{(2)}(\vec{r}_{2}) + E_{\rm vac}^{(2)}(\vec{r}_{1}-\vec{r}_{2}),
\end{equation}
where the first two terms represent the second-order self-energy corrections of each particle, 
\begin{equation}
E_{\rm vac}^{(2)}(\vec{r}_{i}) = -\sum_{E^{(0)}_{n}\neq 0} \frac{\langle 0|H_{{\rm int},i}|E^{(0)}_{n}\rangle \langle E^{(0)}_{n}|H_{{\rm int},i}|0\rangle}{E^{(0)}_{n}},
\end{equation}
while the third term depends on the relative separation of the particles and is the potential energy we are seeking:
\begin{equation}
E_{\rm vac}^{(2)}(\vec{r}_{1}-\vec{r}_{2}) = 
	-\sum_{E^{(0)}_{n}\neq 0}\left[\frac{\langle 0|H_{{\rm int},1}|E^{(0)}_{n}\rangle \langle E^{(0)}_{n}|H_{{\rm int},2}|0\rangle}{E^{(0)}_{n}} + {\rm c.c.}\right],
\label{E vac 2b}
\end{equation}
where ``c.c.'' means complex conjugate, which in this case simply interchanges particles \#1 and \#2.

To evaluate Eq.~(\ref{E vac 2b}), we need to calculate the matrix element $\langle E^{(0)}_{n}|H_{{\rm int},i}|0\rangle$. Since $|E^{(0)}_{n}\rangle \neq |0\rangle$, the only nonzero matrix elements of $\langle E^{(0)}_{n}|H_{{\rm int},i}|0\rangle$ arise for the intermediate state $|E_{n}^{(0)}\rangle =|\vec{k}',s'\rangle_{\nu}|\vec{k},s\rangle_{\bar{\nu}}$ which consists of a neutrino with momentum $\vec{k}'$ and spin $s'$ and an antineutrino with momentum $ \vec{k}$ and spin~$s$ with total energy $E_{n}^{(0)} = \omega_{\vec{k'}} + \omega_{\vec{k}}$.
Using these results, we can rewrite Eq.~(\ref{E vac 2b}) as
\begin{equation}
E_{\rm vac}^{(2)}(\vec{r}_{1}-\vec{r}_{2}) = 
	-\sum_{\vec{k}',\vec{k}} \sum_{s,s'}\left\{
	\frac{\left[\langle 0|H_{{\rm int},1}|\vec{k}',s'\rangle_{\nu}|\vec{k},s\rangle_{\bar{\nu}}\right] 
	\left[_{\bar{\nu}}\langle \vec{k},s| \,_{\nu}\langle \vec{k}',s'|H_{{\rm int},2}|0\rangle\right]}{\omega_{\vec{k'}} + \omega_{\vec{k}}} + {\rm c.c.}\right\}.
\label{E vac 2c}
\end{equation}
The subsequent calculations of Eq.~(\ref{E vac 2c}) are straightforward and described in Appendix~\ref{Calculation Appendix}, giving the two-neutrino exchange potential
\begin{equation}
V_{\nu,\bar{\nu}}(r) = \frac{G_{F}^{2}g_{V,1}^{f}g_{V,2}^{f}m_{\nu}^{3}}{4\pi^{3}r^{2}}K_{3}(2m_{\nu}r).
\end{equation}
To obtain the limit for  massless neutrinos, we use $K_{3}(2x)  \simeq 1/x^{3}$ if  $x \ll 1$, which gives
\begin{equation}
\lim_{m_{\nu}\rightarrow 0} V_{\nu,\bar{\nu}}(r) = \frac{G_{F}^{2}g_{V,1}^{f}g_{V,2}^{f}}{4\pi^{3}r^{5}}.
\end{equation}
All of these results are in agreement with previous work \cite{FS,Hsu,Segarra,Fischbach PRD,Fischbach AoP,Grifols,LP}.  Now that we have demonstrated the efficacy of our approach for deriving the 2NEP for a single neutrino, we will next include the effects of neutrino mixing with three neutrinos.

\section{Interaction Hamiltonians with Mixing}

\subsection{Free Hamiltonian with Three Neutrinos}

Observations reveal there are three flavors of neutrinos, $\nu_{e}$, $\nu_{\mu}$, and $\nu_{\tau}$, while the phenomena of neutrino oscillations indicate that these flavors do not have definite masses \cite{2015 Nobel Prize}.  Let $\nu_{a}(\vec{r})$ represent the quantum field of a neutrino with mass $m_{a}$, where $a = 1,2,3$. The Standard Model fermions interact with neutrinos through flavor fields which are linear combinations of the mass fields  \cite{Shrock PLB,Shrock PRD,Ho},
\begin{eqnarray}
\nu_{\alpha}(\vec{r}) &\equiv& \sum_{a = 1}^{3}U_{\alpha a} \, \nu_{a}(\vec{r}),\\
\label{nu alpha a relation}
\nonumber
\nu_{\alpha}^\dag(\vec{r}) &\equiv& \sum_{a = 1}^{3}U_{\alpha a}^* \, \nu_{a}^\dag(\vec{r}),
\end{eqnarray}
where $\alpha = e, \mu, \tau$, and $U_{\alpha a}$ is  the Pontecorvo-Maki-Nakagawa-Sakata (PMNS) matrix given by
\begin{equation}
U_{\rm PMNS} = 
\left(
\begin{array}{ccc}
U_{e1} & U_{e2} & U_{e3} \\
U_{\mu 1} & U_{\mu 2} & U_{\mu 3} \\
U_{\tau 1} & U_{\tau 2} & U_{\tau 3} 
\end{array}
\right)
=
 \left(
\begin{array}{ccccc}
c_{12}c_{13} && s_{12}c_{13} && s_{13}e^{-i\delta_{CP}} \\
-s_{12}c_{23} - c_{12}s_{23}s_{13}e^{i\delta_{CP}} &&  c_{12}c_{23} - s_{12}s_{23}s_{13}e^{i\delta_{CP}} && s_{23}c_{13} \\
s_{12}s_{23} - c_{12}c_{23}s_{13}e^{i\delta_{CP}} && -c_{12}s_{23}-s_{12}c_{23}s_{13}e^{i\delta_{CP}} && c_{23}c_{13},
\end{array}
\right),
\end{equation}
where $s_{ab} = \sin\theta_{ab}$, $c_{ab} = \cos\theta_{ab}$, and $\delta_{CP}$ is the $CP$-violation phase.  
The mass fields in the Schr\"{o}dinger picture may be written as the generalizations of Eqs.~(\ref{psi}) and (\ref{psi bar}),
\begin{subequations}
\begin{eqnarray}
\nu_{a}(\vec{r}) & = &  \sum_{\vec{k},s}\left(\frac{m_{a}}{V\omega_{\vec{k},a}}\right)^{1/2}
	\left[b_{\vec{k},s,a}u_{s,a}(\vec{k})e^{i\vec{k} \cdot \vec{r}} + d^{\dag}_{\vec{k},s,a}v_{s,a}(\vec{k})e^{-i\vec{k}\cdot \vec{r}}\right], 
\label{psi a} \\
\nu_{a}^{\dag}(\vec{r}) & = &  \sum_{\vec{k},s}\left(\frac{m_{a}}{V\omega_{\vec{k},a}}\right)^{1/2}
	\left[b_{\vec{k},s,a}^{\dag}u^{\dag}_{s,a}(\vec{k})e^{-i\vec{k} \cdot \vec{r}} + d_{\vec{k},s,a}v^{\dag}_{s,a}(\vec{k})e^{i\vec{k}\cdot \vec{r}}\right],
\label{psi a bar}
\end{eqnarray}
\end{subequations}
where $\omega_{\vec{k},a}^{2} = m_{a}^{2} + \vec{k}^{2}$, and the creation and annihilation operators satisfy the usual anticommutation relations where operators associated with different masses commute.  The Hamiltonian can then be written as
\begin{equation}
H_{0} = \sum_{a =1}^{3}\sum_{\vec{k},s}\omega_{\vec{k},a}\left(b_{\vec{k},s,a}^{\dag}b_{\vec{k},s,a} + d_{\vec{k},s,a}^{\dag}d_{\vec{k},s,a}\right),
\label{H free}
\end{equation}
where the infinite vacuum energy has been dropped as before.

It is important to note that the vacuum that we will use subsequently in perturbation theory calculations is the eigenstate of the Hamiltonian in Eq.~(\ref{H free}), which is simply given by the  product of the vacuum states of each neutrino mass field, 
\begin{equation}
|0\rangle \rightarrow \prod_{a=1}^{3}|0\rangle_{a}.
\label{neutrino vacuum}
\end{equation} 
For simplicity, we will write this vacuum state as $|0\rangle$ in our calculations.  It has been noted that there are a variety of theoretical difficulties associated with using this vacuum \cite{BV PLB,BV weak interaction decay}, and that other constructions of the neutrino vacuum provide ways to circumvent these issues \cite{BV AoP, BV PLB}. Meanwhile, there are other proposals that resolve these theoretical difficulties without resorting to these alternative neutrino vacua \cite{Ho,Lobanov}. Furthermore, it can be shown that the results in Section~\ref{Potential Section} will persist as a contribution in these models, though the detailed discussion is beyond the scope of this paper. Generally, it is clear that the effects of neutrino mixing are a direct consequence of the bilinear form of the neutrino field in the weak interaction Hamiltonian.

\subsection{Interaction Hamiltonians with Three Neutrino Flavors}

\subsubsection{Flavor Fields}

 According to the Standard Model, the interaction  of neutrinos with other fermions is through the flavor fields $\nu_{\alpha}(\vec{r})$, not the mass fields $\nu_{a}(\vec{r})$.  In this section, we will describe the low-energy effective Hamiltonians describing the interaction of the three flavors of neutrinos interacting with protons, neutrons, and charged leptons.  We will not explore the interactions involving individual quarks which are bound in baryons and mesons.   In our low-energy theory, the nucleons will be  treated effectively as fundamental particles.

\begin{figure}[tbp]
\begin{center}
\includegraphics[width=3in]{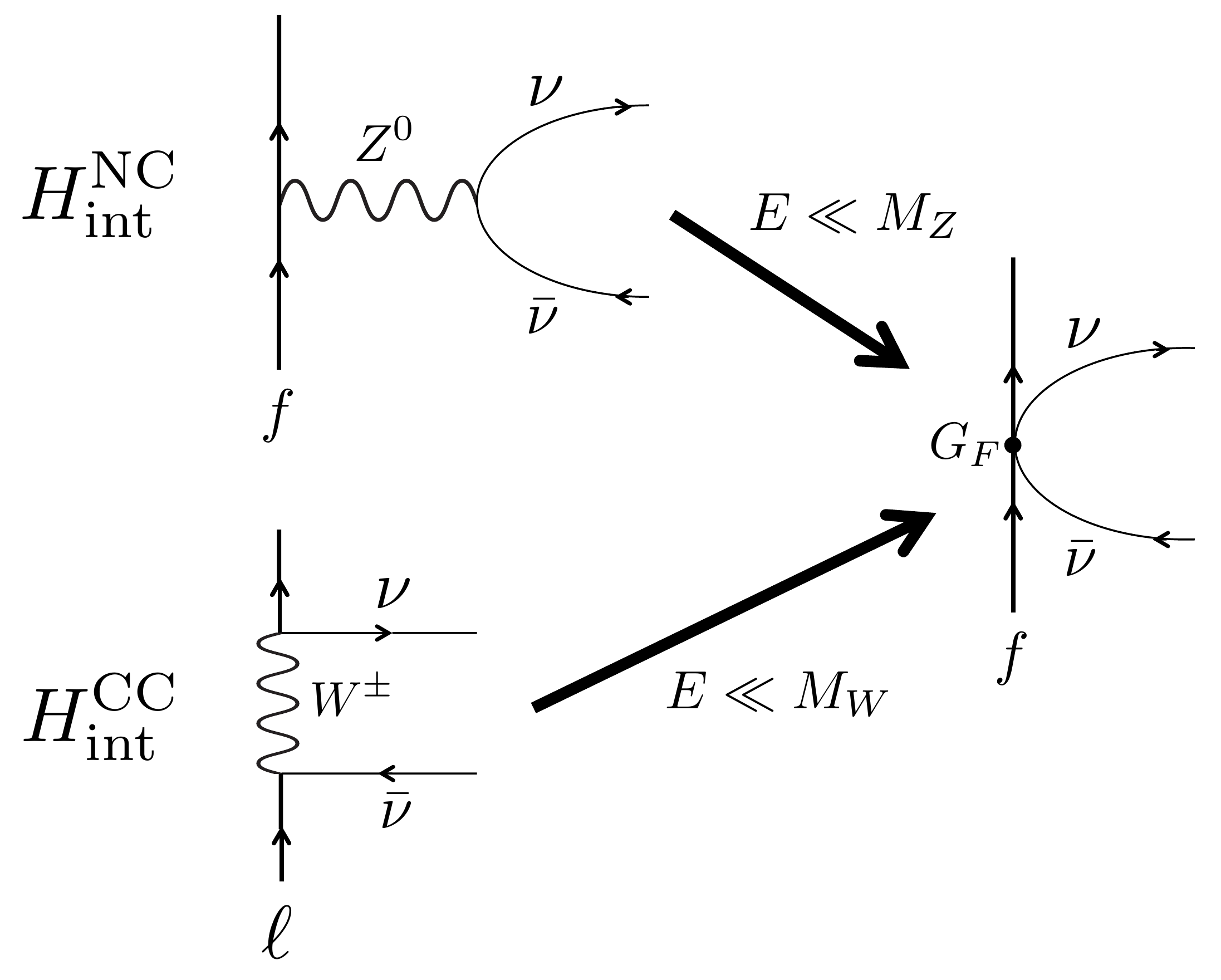}
\caption{Feynman diagrams showing how the neutral current (NC) (top) and  charged current (CC) (bottom) processes involving vector bosons $Z^{0}$ and $W^{\pm}$ and neutrinos lead to the effective process described by Eqs.~(\ref{H int NC}) and (\ref{H int CC}) when the energies involved satisfy $E \ll M_{Z}, M_{W}$.   Note that nucleons and charged leptons interact via NC processes while only leptons participate in the CC processes.}
\label{NC-CC diagrams}
\end{center}
\end{figure}

  In general, the low-energy effective Hamiltonian density describing the interaction of  neutrinos with  fermions is the sum of two contributions \cite{Giunti book},
\begin{equation}
{\cal H}_{\rm int}(\vec{r}) = {\cal H}_{\rm int}^{\rm NC}(\vec{r}) + {\cal H}_{\rm int}^{\rm CC}(\vec{r}),
\end{equation}
where the neutral current (NC) Hamiltonian density describing weak interactions between neutrinos and fermions is
\begin{equation}
{\cal H}_{\rm int}^{\rm NC}(\vec{r}) = \frac{G_{F}}{\sqrt{2}}\left[\sum_{f}\bar{f}(\vec{r})\gamma_{\sigma}\left(g_{V}^{f} - g_{A}^{f}\gamma^{5}\right)f(\vec{r})\right]\left[\sum_{\alpha = e,\mu,\tau} \bar{\nu}_{\alpha}(\vec{r})\gamma^{\sigma}\left(1 - \gamma^{5}\right)\nu_{\alpha}(\vec{r})\right].
\label{H int NC}
\end{equation}
The charged current (CC) interaction Hamiltonian density ${\cal H}_{\rm int}^{\rm CC}(\vec{r})$ only involves charged leptons interacting with their corresponding flavors of neutrinos:
\begin{equation}
{\cal H}_{\rm int}^{\rm CC}(\vec{r}) = \frac{G_{F}}{\sqrt{2}}\left[\bar{\ell}_{\alpha}(\vec{r})\gamma_{\sigma}\left(1 - \gamma^{5}\right)\ell_{\alpha}(\vec{r})\right]\left[ \bar{\nu}_{\alpha}(\vec{r})\gamma^{\sigma}\left(1 - \gamma^{5}\right)\nu_{\alpha}(\vec{r})\right],
\label{H int CC}
\end{equation}
where $\ell_{\alpha}(\vec{r})$ is the charged lepton field with flavor $\alpha$.  Fig.~\ref{NC-CC diagrams} shows the Feynman diagrams which illustrate how the more fundamental Standard Model NC and CC processes involving the vector bosons $Z^{0}$  and $W^{\pm}$ with masses $M_{Z}$ and $M_{W}$ reduce to the lower energy interactions involving just fermions and neutrino propagators when the energies satisfy $E \ll M_{Z},M_{W}$.

Since we are interested in only neutrino interactions with nonrelativistic fermions, we focus on only the spin-independent interaction Hamiltonians involving nucleons and charged leptons.  Nucleons only experience the NC interaction so their interaction Hamiltonian is the generalization of Eq.~(\ref{H int i}),
\begin{equation}
H_{\rm int, N}(\vec{r}_{i}) =\frac{G_{F}g_{V}^{N}}{\sqrt{2}}\left[\sum_{\alpha = e,\mu,\tau} \nu_{\alpha}^{\dag}(\vec{r}_{i})\left(1 - \gamma^{5}\right)\nu_{\alpha}(\vec{r}_{i})\right],
\label{H N flavor}
\end{equation}
where N = p,n (protons, neutrons).  On the other hand, the charged lepton interaction Hamiltonian includes NC and CC contributions, and is given by
\begin{equation}
H_{\rm int, \alpha}(\vec{r}_{i}) = \frac{G_{F}}{\sqrt{2}}\left\{ g_{V}^{\alpha}\left[\sum_{\beta = e,\mu,\tau} \nu_{\beta}^{\dag}(\vec{r}_{i})\left(1 - \gamma^{5}\right)\nu_{\beta}(\vec{r}_{i})\right] + \nu_{\alpha}^{\dag}(\vec{r}_{i})\left(1 - \gamma^{5}\right)\nu_{\alpha}(\vec{r}_{i}) \right\}.
\label{H lepton flavor}
\end{equation}

\subsubsection{Mass Fields}

\paragraph{Nucleons.}

While the interaction between neutrinos and other fermions is most naturally expressed in terms of the flavor fields $\nu_{\alpha}(\vec{r})$, we need to express the interaction in terms of the mass fields $\nu_{a}(\vec{r})$ to calculate the two-neutrino exchange potential in our formalism.  Using Eq.~(\ref{nu alpha a relation}), it is straightforward to show that the neutral current interaction is independent of flavor so 
\begin{equation}
\sum_{\alpha = e,\mu,\tau} \nu_{\alpha}^{\dag}(\vec{r})\left(1 - \gamma^{5}\right)\nu_{\alpha}(\vec{r}) = \sum_{a = 1}^{3} \nu_{a}^{\dag}(\vec{r})\left(1 - \gamma^{5}\right)\nu_{a}(\vec{r}).
\label{NC field current transformation}
\end{equation}
Thus, nucleons couple equally to the three types of neutrino fields so Eq.~(\ref{H N flavor}) is easily rewritten in terms of the neutrino mass fields:
\begin{equation}
H_{\rm int, N}(\vec{r}_{i}) =\frac{G_{F}g_{V}^{N}}{\sqrt{2}}\left[\sum_{a = 1}^{3} \nu_{a}^{\dag}(\vec{r}_{i})\left(1 - \gamma^{5}\right)\nu_{a}(\vec{r}_{i})\right].
\label{H int N}
\end{equation}

\paragraph{Charged Leptons.}  

The interaction of the neutrino fields with the charged lepton is a bit more complicated due to the additional contribution from the charged current interaction.  Transforming the flavor neutrino field in the neutrino current into mass fields gives
\begin{eqnarray}
\nu_{\alpha}^{\dag}(\vec{r})\left(1 - \gamma^{5}\right)\nu_{\alpha}(\vec{r})  
	& = &   \sum_{a,b  = 1}^{3} U^{*}_{\alpha a}U_{\alpha b} \left[ \nu_{a}^{\dag}(\vec{r})\left(1 - \gamma^{5}\right) \nu_{b}(\vec{r})\right]
\nonumber  \\
	& = &   \sum_{a = 1}^{3} |U_{\alpha a}|^{2} \left[ \nu_{a}^{\dag}(\vec{r})\left(1 - \gamma^{5}\right) \nu_{a}(\vec{r})\right] 
		+ \sum_{a \neq b} U^{*}_{\alpha a}U_{\alpha b} \left[ \nu_{a}^{\dag}(\vec{r})\left(1 - \gamma^{5}\right) \nu_{b}(\vec{r})\right].
\label{CC field current transformation}
\end{eqnarray}
If we now substitute Eqs.~(\ref{NC field current transformation}) and (\ref{CC field current transformation}) into Eq.~(\ref{H lepton flavor}), we  obtain the  interaction Hamiltonian for a charged lepton located at position $\vec{r}_{i}$,
\begin{equation}
H_{\rm int, \alpha}(\vec{r}_{i}) = 
	 \frac{G_{F}}{\sqrt{2}}\left\{\left[\sum_{a = 1}^{3} \left(g_{V}^{\alpha} +|U_{\alpha a}|^{2} \right) \nu_{a}^{\dag}(\vec{r}_{i})\left(1 - \gamma^{5}\right)\nu_{a}(\vec{r}_{i})\right]  + \sum_{a \neq b} U^{*}_{\alpha a}U_{\alpha b} \left[ \nu_{a}^{\dag}(\vec{r}_{i})\left(1 - \gamma^{5}\right) \nu_{b}(\vec{r}_{i})\right] \right\}.
\label{H int lepton}
\end{equation}
We see that $H_{\rm int, \alpha}(\vec{r}_{i})$ naturally divides into the sum of  terms of neutrino currents involving the same and different mass neutrino fields:
\begin{equation}
H_{\rm int, \alpha}(\vec{r}_{i}) = \sum_{a = 1}^{3}H_{\rm int, \alpha}^{(aa)}(\vec{r}_{i})  + \sum_{a \neq b}H_{\rm int, \alpha}^{(ab)}(\vec{r}_{i}),
\label{H int lepton split}
\end{equation}
where
\begin{equation}
H_{\rm int, \alpha}^{(aa)}(\vec{r}_{i}) \equiv	 \frac{G_{F}}{\sqrt{2}} \left[ \left(g_{V}^{\alpha} +|U_{\alpha a}|^{2} \right) \nu_{a}^{\dag}(\vec{r}_{i})\left(1 - \gamma^{5}\right)\nu_{a}(\vec{r}_{i})\right]
\label{H lepton aa}
\end{equation}
and
\begin{equation}
H_{\rm int, \alpha}^{(ab)}(\vec{r}_{i}) \equiv	 \frac{G_{F}}{\sqrt{2}}\,\left\{U^{*}_{\alpha a}U_{\alpha b} \left[ \nu_{a}^{\dag}(\vec{r}_{i})\left(1 - \gamma^{5}\right) \nu_{b}(\vec{r}_{i}) \right] \right\}.
\label{H lepton ab}
\end{equation}
This division will result in two distinct ways in which neutrino mixing will affect the 2NEP.  In Eq.~(\ref{H lepton aa}), we see that the mixing in $H_{\rm int, \alpha}^{(aa)}(\vec{r}_{i})$ results only in a change of the coefficient $g_{V}^{\alpha} \rightarrow (g_{V}^{\alpha} +|U_{\alpha a}|^{2})$ from the single neutrino result and will not substantively affect the spatial dependence of the 2NEP.  On the other hand, $H_{\rm int, \alpha}^{(ab)}(\vec{r}_{i})$ given by Eq.~(\ref{H lepton ab}) will result in a modified spatial dependence of the 2NEP between leptons due to the interference of different mass neutrino contributions.

\section{Interaction Potentials with Mixing}
\label{Potential Section}

\subsection{Overview}

In this section, we will use the Hamiltonians describing the interactions of fermions with the neutrino fields given in the previous section to derive the 2NEPs for nucleon-nucleon, nucleon-lepton, and lepton-lepton interactions including neutrino mixing.  As in the single neutrino case, the derivation of the 2NEP with mixing will start with the second-order energy shift of the vacuum due to two fermions \#1 and \#2 which depends on the fermions separation, Eq.~(\ref{E vac 2b}),
\begin{equation}
E_{\rm vac}^{(2)}(\vec{r}_{1}-\vec{r}_{2}) = 
	-\sum_{E^{(0)}_{n}\neq 0}\left[\frac{\langle 0|H_{{\rm int},1}|E^{(0)}_{n}\rangle \langle E^{(0)}_{n}|H_{{\rm int},2}|0\rangle}{E^{(0)}_{n}} + {\rm c.c.}\right].
\label{E vac 2b again}
\end{equation}
where $H_{{\rm int},i}$ now will become $H_{\rm int, N}(\vec{r}_{i})$, given by Eq.~(\ref{H int N}) or $H_{\rm int, \alpha}(\vec{r}_{i})$ given by Eqs.~(\ref{H int lepton split})--(\ref{H lepton ab}), depending upon the identities of the interacting fermions.

\subsection{Potentials for Two Nucleons}

Since the interaction of neutrinos with nucleons is flavor independent, the derivation of the 2NEP is a straightforward extension of the single neutrino case.  For two nucleons, the second-order vacuum energy shift is obtained from Eq.~(\ref{E vac 2b}) with the replacement $H_{{\rm int},i} \rightarrow H_{{\rm int,N}_{i}, i}$, ${\rm N}_{i} = {\rm p,n}$ given by Eq.~(\ref{H int N}):
\begin{equation}
E_{\rm vac,NN}^{(2)}(\vec{r}_{1}-\vec{r}_{2}) = 
	-\sum_{E^{(0)}_{n} \neq 0}\left\{
	\frac{\langle 0| H_{{\rm int,N}_{1}, 1}|E^{(0)}_{n} \rangle
	\langle E^{(0)}_{n} | H_{{\rm int,N}_{2}, 2} |0\rangle}{E^{(0)}_{n} } + {\rm c.c.}\right\}.
\label{E vac 2 nucleons a}
\end{equation}
The only intermediate states that will give a nonzero contribution are neutrino-antineutrino pairs of the same mass state $a$, $|\vec{k}',s'\rangle_{\nu_{a}}|\vec{k},s\rangle_{\bar{\nu}_{a}}$, 
which gives
\begin{equation}
E_{\rm vac,NN}^{(2)}(\vec{r}_{1}-\vec{r}_{2}) = -\sum_{a =1}^{3}\sum_{\vec{k}',\vec{k}} \sum_{s,s'}\left\{
	\frac{\left[\langle 0|H_{{\rm int,N}_{1}, 1}|\vec{k}',s'\rangle_{\nu_{a}}|\vec{k},s\rangle_{\bar{\nu}_{a}}\right] 
	\left[_{\bar{\nu}_{a}}\langle \vec{k},s| \,_{\nu_{a}}\langle \vec{k}',s'| H_{{\rm int,N}_{2}, 2} |0\rangle\right]}{\omega_{\vec{k'}} + \omega_{\vec{k}}} + {\rm c.c.}\right\}.
\label{E vac 2 nucleons b}
\end{equation}
Graphically, the process leading to the nucleon-nucleon 2NEP is shown in Fig.~\ref{2NEP diagram}(a).
\begin{figure}[t]
\begin{center}
\includegraphics[width=6.5in]{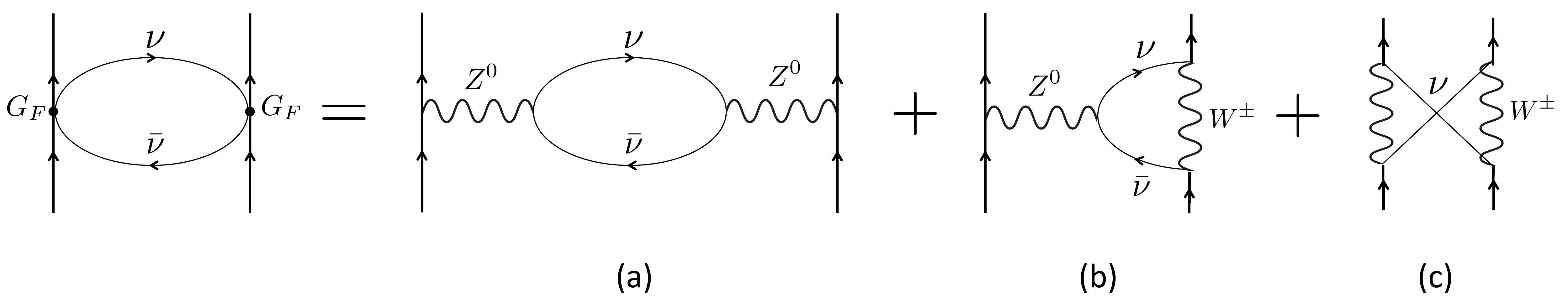}
\caption{Feynman diagrams contributing to the 2NEP.   Here (a) represents the NC-NC interaction, (b) one of the two NC-CC diagrams contributing to the lepton weak form factor, and (c) the CC-CC diagram.  For the nucleon-nucleon 2NEP, only the diagram (a) contributes, while diagrams (a) and (b) contribute to the nucleon-lepton 2NEP.  Finally, (a), (b), and (c) contribute to the lepton-lepton 2NEP.}
\label{2NEP diagram 2}
\end{center}
\end{figure}
The interaction energy between any two nucleons is then the sum of single neutrino potentials of each of the three mass states:
\begin{equation}
V_{\rm N_{1},N_{2}}(r) = \frac{G_{F}^{2}g_{V,1}^{{\rm N}_{1}}g_{V,2}^{{\rm N}_{2}}}{4\pi^{3}r^{2}}\sum_{a = 1}^{3}m_{a}^{3}K_{3}(2m_{a}r).
\label{general nucleon V}
\end{equation}
In the limit $r \ll m_{a}^{-1}$ for all $a$, this reduces to
\begin{equation}
 V_{\rm N_{1},N_{2}}(r) \simeq  \frac{3G_{F}^{2}g_{V,1}^{{\rm N}_{1}}g_{V,2}^{{\rm N}_{2}}}{4\pi^{3}r^{5}},
\label{massless nucleon V}
\end{equation}
which is three times the single neutrino result.  The nucleon-nucleon 2NEP in this separation regime is proprotional to the total number of neutrinos since all contribute equally in the virtual exchange.

\subsection{Potentials for  Nucleon-Lepton Interaction}

\begin{figure}[t]
	\centering
	\includegraphics[width=8cm]{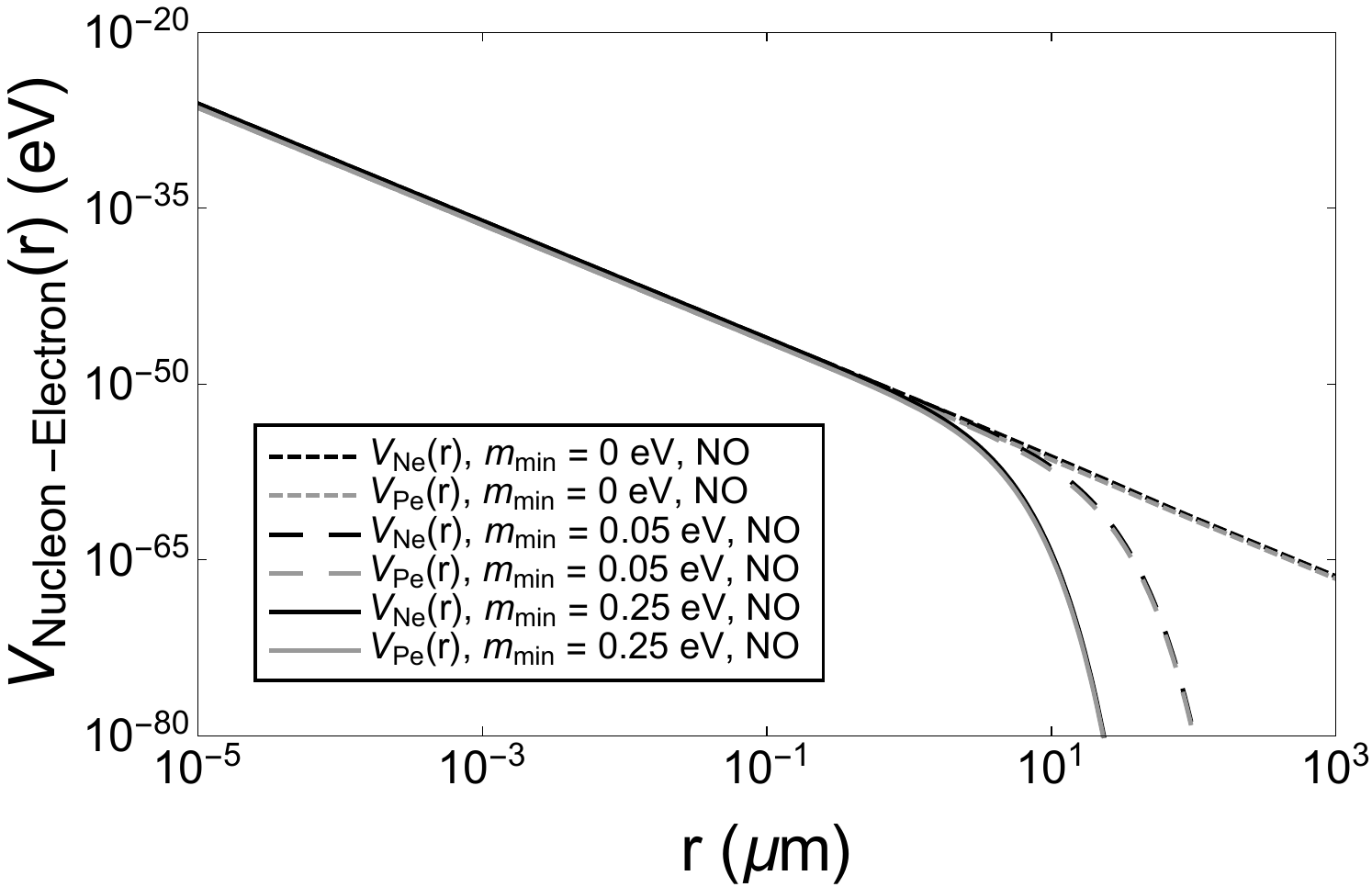}	
	\caption{Plots of the total 2NEP between nucleons and electrons with normal ordering (NO) of neutrino masses: the black lines represent the neutron-electron 2NEP, while the gray lines represent the proton-electron 2NEP. We assumed the smallest neutrino mass state  $m_{\rm min}=0 \ \textrm{eV}$ (short-dashed), $m_{\rm min}=0.05\ \textrm{eV}$ (long-dashed) and $m_{\rm min}=0.25\ \textrm{eV}$ (solid). Values for all other parameters used were obtained from the Particle Data Group \cite{PDG 2018} given in Table~\ref{PMNS table}.}
	\label{fig potential nucleon electron plot}
\end{figure}

Now let us consider the interaction energy for a nucleon (particle \#1) and a charged lepton (particle \#2).  Then the second-order vacuum shift is 
\begin{equation}
E_{{\rm vac,N}\alpha}^{(2)}(\vec{r}_{1}-\vec{r}_{2}) = 
	-\sum_{E^{(0)}_{n} \neq 0}\left\{
	\frac{\langle 0| H_{{\rm int,N}, 1}|E^{(0)}_{n} \rangle
	\langle E^{(0)}_{n} | H_{{\rm int,}\alpha, 2} |0\rangle}{E^{(0)}_{n} } + {\rm c.c.}\right\},
\label{E vac 2 nucleon-electron a}
\end{equation}
where N = p, n.  Here $H_{{\rm int,N}, 1}$ is given by Eq.~(\ref{H int N}) while $H_{{\rm int}, \alpha, 2}$ is given by Eq.~(\ref{H int lepton}).  Like the nucleon-nucleon case, the only nonzero contributions will arise when the intermediate states are neutrino-antineutrino pairs of the same mass state $a$, $|\vec{k}',s'\rangle_{\nu_{a}}|\vec{k},s\rangle_{\bar{\nu}_{a}}$, 
\begin{equation}
E_{{\rm vac,N}\alpha}^{(2)}(\vec{r}_{1}-\vec{r}_{2}) = -\sum_{a =1}^{3}\sum_{\vec{k}',\vec{k}} \sum_{s,s'}\left\{
	\frac{\left[\langle 0|H_{{\rm int,N}, 1}|\vec{k}',s'\rangle_{\nu_{a}}|\vec{k},s\rangle_{\bar{\nu}_{a}}\right] 
	\left[_{\bar{\nu}_{a}}\langle \vec{k},s| \,_{\nu_{a}}\langle \vec{k}',s'| H_{{\rm int}, \alpha,2}^{(aa)} |0\rangle\right]}{\omega_{\vec{k'}} + \omega_{\vec{k}}} + {\rm c.c.}\right\}.
\label{E vac 2 nucleon-electron b}
\end{equation}
 Graphically, the two processes contributing to the nucleon-lepton 2NEP are shown in Fig.~\ref{2NEP diagram 2}.
Like the nucleon-nucleon potential,  the interaction energy between a nucleon and an electron is the sum of single neutrino potentials of each of the three mass states, but now incorporates an additional factor which depends on mixing which arises from the NC-CC diagram (b) in Fig.~\ref{2NEP diagram 2}:
\begin{equation}
V_{{\rm N}\alpha}(r) = \frac{G_{F}^{2}g_{V}^{\rm N}}{4\pi^{3}r^{2}}\sum_{a = 1}^{3}m_{a}^{3} \left(g_{V}^{\alpha} +|U_{\alpha a}|^{2} \right) K_{3}(2m_{a}r).
\label{general nucleon lepton V}
\end{equation}
When $r \ll m_{a}^{-1}$ for all $a$, the nucleon-lepton 2NEP reduces
\begin{equation}
V_{\rm N\alpha}(r) \simeq \frac{G_{F}^{2}g_{V}^{\rm N}}{4\pi^{3}r^{5}} \left(3g_{V}^{e} + \sum_{a = 1}^{3}|U_{\alpha a}|^{2}\right) = \frac{G_{F}^{2}g_{V}^{\rm N}}{4\pi^{3}r^{5}} \left(3g_{V}^{e} + 1\right),
\label{massless nucleon electron V}
\end{equation}
where in the last step we used the universal neutral current coupling to leptons ($g_{V}^{\alpha} = g_{V}^{e}$) and the general property of a unitary matrix that its rows form an orthonormal basis \cite{Rasin}.
Unlike the nucleon-nucleon case, we see that potential between a nucleon and a lepton  depends on mixing, even in this limit, but the mixing does not qualitatively alter the spatial dependence.  In the massless limit, the NC and CC current interactions effectively make three and one contributions, respectively, to the nucleon-electron 2NEP.  Graphs of the  nucleon-electron 2NEP assuming normal ordering of neutrino masses are shown in Fig.~\ref{fig potential nucleon electron plot}.

\begin{table}[t]
\caption{Neutrino mass and PMNS matrix parameters used in numerical calculations (Table~14.1, Ref.~\cite{PDG 2018}).  Here normal ordering assumes $m_{1} < m_{2} < m_{3}$, while inverted ordering assumes $m_{3} < m_{1} < m_{2}$.  For the graphs, we assumed three possible values for the smallest neutrino mass state: $m_{\rm min} = 0~{\rm eV}, 0.05~{\rm eV}$, and 0.25~eV.}
\begin{center}
\begin{tabular}{lcccc} \hline\hline
Parameter && Normal Ordering (NO) && Inverted Ordering (IO) \\ \hline
$\Delta m_{21}^{2}$ [$10^{-5}$~eV$^{2}$] && \multicolumn{3}{c}{7.37} \\
$\Delta m_{31 (23)}^{2}$ [$10^{-3}$~eV$^{2}$] && 2.56 && 2.54 \\
$\sin^{2}\theta_{12}$ && \multicolumn{3}{c}{0.297} \\
$\sin^{2}\theta_{13}$ && $0.0215$ && $0.0216$ \\
$\sin^{2}\theta_{23}$ && $0.425$ && $0.589$ \\
$\delta_{CP}$ && $1.38\pi$ && $1.31\pi$  \\ \hline\hline
\end{tabular}
\end{center}
\label{PMNS table}
\end{table}%

\subsection{Potentials for  Two Electrons}

Now let us consider the interaction energy between two electrons. The three diagrams contributing to the general lepton-lepton 2NEP are shown in Fig.~\ref{2NEP diagram 2}. The second-order vacuum shift is 
\begin{equation}
E_{\rm vac,ee}^{(2)}(\vec{r}_{1}-\vec{r}_{2}) = 
-\sum_{E^{(0)}_{n} \neq 0}\left\{
\frac{\langle 0| H_{{\rm int,e}, 1}|E^{(0)}_{n} \rangle
	\langle E^{(0)}_{n} | H_{{\rm int, e}, 2} |0\rangle}{E^{(0)}_{n} } + {\rm c.c.}\right\},
\label{E vac 2 electron-electron a}
\end{equation}
where $H_{{\rm int,e}, i}$ is given by Eq.~(\ref{H int lepton split}) with $\alpha = {\rm e}$.   From Eq.~(\ref{H int lepton split}), we see that Eq.~(\ref{E vac 2 electron-electron a}) can be grouped into 2 separate contributions,
\begin{equation}
E_{\rm vac,ee}^{(2)}(\vec{r}_{1}-\vec{r}_{2}) = E_{\rm vac,ee}^{(2)(a=b)}(\vec{r}_{1}-\vec{r}_{2}) + E_{\rm vac,ee}^{(2)(a\neq b)}(\vec{r}_{1}-\vec{r}_{2}).
\label{E vac 2 electron-electron split}
\end{equation}
The first term  $ E_{\rm vac,ee}^{(2)(a=b)}(\vec{r}_{1}-\vec{r}_{2})$ arises from Eq.~(\ref{H lepton aa}), which corresponds to the case of exchanging a neutrino-antineutrino pair of the same mass state similar to the nucleon-nucleon and nucleon-lepton case. The second contribution in Eq.~(\ref{E vac 2 electron-electron split}), $E_{\rm vac,ee}^{(2)(a\neq b)}(\vec{r}_{1}-\vec{r}_{2})$ comes  from Eq.~(\ref{H lepton ab}), which corresponds to the case of exchanging a neutrino from one mass state with an antineutrino from another mass state.  [Because of the difference in the virtual neutrinos exchanged, there is no interference term involving both Eqs.~(\ref{H lepton aa}) and (\ref{H lepton ab}).]  Writing out explicitly this new contribution due to mixing, we find
\begin{equation}
\begin{split}
E_{\rm vac,ee}^{(2)(\textrm{mix})}(\vec{r}_{1}-\vec{r}_{2}) \equiv  E_{\rm vac,ee}^{(2)(a\neq b)}(\vec{r}_{1}-\vec{r}_{2}) = &
-\sum_{E^{(0)}_{n} \neq 0}\frac{1}{E_{n}^{(0)}}\left\{
\left[\langle 0|  \sum_{a \neq b} \frac{G_F U^{*}_{ea}U_{eb}^{}}{\sqrt{2}} \nu_{a}^{\dag}(\vec{r}_{1})  \left(1 - \gamma^{5}\right) \nu_{b}(\vec{r}_{1}) |E^{(0)}_{n} \rangle\right] \right.
\\
&
\left. \times \left[\langle E^{(0)}_{n} |  \sum_{a' \neq b'} \frac{ G_F U^{*}_{ea'}U_{eb'}^{}}{\sqrt{2}}   \nu_{a'}^{\dag}(\vec{r}_{2})\left(1 - \gamma^{5}\right) \nu_{b'}(\vec{r}_{2})  |0\rangle\right]
+ \left(1 \leftrightarrow  2\right)
\right\}.
\end{split}
\label{E vac 2 electron electron mix a}
\end{equation}
The nonzero terms in this quadruple sum occur only when $a'=b$ and $b'=a$ due to the matching of exchanged particles, so the final contribution from mixing is given by
\begin{equation}
E_{\rm vac,ee}^{(2)(\textrm{mix})}(\vec{r}_{1}-\vec{r}_{2}) = -\sum_{a > b} G_F^2 |U_{ea}|^2 |U_{eb}|^2 \sum_{E^{(0)}_{n} \neq 0}\left\{
\frac{\langle 0| \nu_{a}^{\dag}(\vec{r}_{1})\left(1 - \gamma^{5}\right) \nu_{b}(\vec{r}_{1}) |E^{(0)}_{n} \rangle
\langle E^{(0)}_{n} | \nu_{b}^{\dag}(\vec{r}_{2})\left(1 - \gamma^{5}\right) \nu_{a}(\vec{r}_{2})  |0\rangle}{E^{(0)}_{n} } + \left(1 \leftrightarrow  2\right)\right\}.
\label{E vac 2 electron electron mix b}
\end{equation}
While the contribution  $ E_{\rm vac,ee}^{(2)(a=b)}(\vec{r}_{1}-\vec{r}_{2})$ from  Eq.~(\ref{H lepton aa}) can be evaluated exactly as in the nucleon-nucleon and nucleon-lepton cases, we have not found a closed form expression for $E_{\rm vac,ee}^{(2)(\textrm{mix})}(\vec{r}_{1}-\vec{r}_{2}) $ given by  Eq.~(\ref{E vac 2 electron electron mix b}).  Instead, one can make an expansion in powers of $\left(m_-^{ab}/m_+^{ab} \right)^{2n}$, where
\begin{equation}
m_{\pm}^{ab}  \equiv  m_a \pm m_b,
\end{equation}
which can be evaluated.  To $\mathcal{O} \left[ \left(m_-^{ab}/m_+^{ab}\right)^2 \right]$, the resulting 2NEP between two electrons is given by
\begin{equation}
V_{\rm ee}(r) =  \frac{G_{F}^{2}}{4\pi^{3}r^{2}} \left[
\sum_{a = 1}^{3}m_{a}^{3} \left(g_{V}^{e} +|U_{ea}|^{2} \right)^{2} K_{3}(2m_{a}r)\right]  + V_{\rm ee,mix}(r),
\label{E vac 2 electron-electron b massive}
\end{equation}
where the new contribution due to mixing is
\begin{equation}
V_{\rm ee,mix}(r) =  \frac{G_{F}^{2}}{4\pi^{3}r^{2}}
 \sum_{a > b}^{3}  \frac{ |U_{ea}|^2 |U_{eb}|^2  }{4}
\left\{
m_+^{ab} \left[\left(m_+^{ab}\right)^2 + \left(m_-^{ab}\right)^2 \right] K_3\left(\left.m_+^{ab}\right. r\right)
 - \frac{4 \left(m_-^{ab}\right)^2 }{r} K_2\left(m_+^{ab}\, r\right)
 + \mathcal{O} \left[ \left(\frac{m_-^{ab}}{m_+^{ab}} \right)^2 \right]
\right\}.
\label{E vac 2 electron-electron mix massive}
\end{equation}
 Using the parameters given in Table~\ref{PMNS table}, one finds that the lowest order contribution given by Eq.~(\ref{E vac 2 electron-electron mix massive}) is remarkably accurate, with higher order terms contributing  significantly less than 1\%.
The mixing potential Eq.~(\ref{E vac 2 electron-electron mix massive}) is always repulsive even though it contains attractive and repulsive terms. This result follows because $\left(m_+^{ab}\right)^2 + \left(m_-^{ab}\right)^2 \ge 2\left(m_-^{ab}\right)^2$ and $ K_3(x) > 2 K_2(x)/x$.
When $r \ll m_{a}^{-1}$ for all $a$, we find
\begin{equation}
V_{\rm ee}(r)  \simeq  \frac{G_{F}^{2}}{4\pi^{3}r^{5}}\left[\sum_{a = 1}^{3} \left(g_{V}^{e} +|U_{ea}|^{2} \right)^{2} + 2 \sum_{a>b}^{3} |U_{ea}|^2 |U_{eb}|^2\right].
\label{V ee}
\end{equation}
Using the properties of the PMNS matrix and the universal neutral current coupling to charged leptons, Eq.~(\ref{V ee}) simplifies to
\begin{equation}
V_{\rm ee}(r)  = V_{\mu \mu}(r)  = V_{\tau\tau}(r) \simeq  \frac{G_{F}^{2}}{4\pi^{3}r^{5}}\left[3\left(g_{V}^{e}\right)^{2} + 2 g_{V}^{e} + 1\right].
\end{equation}
This follows because of the flavor independence of the interaction  in the high momentum (small $r$) limit, so there are three contributions from  the NC diagram, Fig~\ref{2NEP diagram 2}(a), two contributions from the NC-CC diagram, Fig~\ref{2NEP diagram 2}(b), and one contribution from  the CC diagram, Fig~\ref{2NEP diagram 2}(c).

It is important to note that the asymptotic expansion used to obtain Eq.~(\ref{E vac 2 electron-electron mix massive}) is only valid for the cases of mixing between 2 massive neutrinos or 2 massless neutrinos.  It fails for the case of mixing between a massless neutrino and a massive neutrino. 
Unlike the case when all three neutrinos are massive, the  contribution to the electron-electron 2NEP from mixing when a single neutrino is massless (here  the $a$th neutrino) is obtained exactly  as
\begin{equation}
\begin{split}
V_{\rm {ee, mix}}^{m_a=0}(r) = &  \frac{ G_{F} ^2} {4 \pi^3 r^5} \sum_{\substack{b =1 \\ b \neq a}}^{3}
	\frac{|U_{ea}|^2 |U_{eb}|^2 } {12} \left[ 
	 e^{-m_b r} \left( 24 + 24 m_b r + 6 m_b^2 r^2 -2  m_b^3 r^3 +  m_b^4 r^4 - m_b^5 r^5 \right) \right.\\
	& \left. -  (6 m_b^4 r^4 + m_b^6 r^6) \ \mathrm{Ei}\left(-m_b r\right) - 6 m_b^4 r^4 \ \Gamma\left( 0, m_b r \right) \right],
\end{split}   
\label{E vac 2 electron-electron b massless}
\end{equation}
where $\mathrm{Ei}(x)$ is the exponential integral Ei and $\Gamma(s,x)$ is the upper incomplete gamma function.

Let us now compare our results for the 2NEP between two electrons with mixing with the integral expression derived by Lusignoli and Petrarca \cite{LP},
\begin{equation}
\begin{split}
V_{\textrm{ee}}^{\rm LP}(r) =  & \frac{G_F^2}{24 \pi^3 r^5} \sum_{a,b = 1}^{3} |U_{ea}|^2 |U_{eb}|^2 
\\
& \times \int_{m_+^{ab} r}^{\infty} 
\sqrt{y^4 - \left[ \left(m_+^{ab}\right)^2 + \left( m_-^{ab} \right)^2 \right] r^2 y^2 + \left( m_+^{ab}\right)^2 \left( m_-^{ab} \right)^2 r^4}
\\
& \times 
\left\{ y^2 - \frac{\left[ \left(m_+^{ab}\right)^2 + \left( m_-^{ab} \right)^2 \right]r^2}{4} + \frac{\left( m_+^{ab}\right)^2 \left( m_-^{ab} \right)^2 r^4}{2y^2} \right\} 
\frac{e^{-y}}{y} dy,
\end{split}  
\label{LP result}
\end{equation}
which we have adapted to our notation.  One can verify that this  result indeed agrees with our mixing contribution $V_{\rm ee,mix}(r)$ given by  Eq.~(\ref{E vac 2 electron-electron mix massive})  by applying the substitution $y = ar$, where  here $a$ is given by 
\begin{equation}
a = \sqrt{m_a^2 + m_b^2 + 2 m_a m_b \cosh t}.
\label{mixing a}
\end{equation}
However, the Lusignoli and Petrarca result Eq.~(\ref{LP result}) does not include the weak NC interaction and its interference with the CC interaction which arises from the diagram in Fig.~\ref{2NEP diagram 2}(b). This additional contribution and interference from the NC results in a modification of the coupling of the electron when $a = b$  from $|U_{ea}|^2 |U_{ea}|^2$ in Eq.~(\ref{LP result}) to $\left(g_{V}^{\rm e} +|U_{{\rm e} a}|^{2} \right)\left(g_{V}^{\rm e} +|U_{{\rm e} a}|^{2} \right) $ in our result.

\subsection{Potentials for Two Leptons}

A straightforward generalization of the calculation carried out in Eq.~(\ref{E vac 2 electron-electron a}) yields the analogous potential to Eqs.~(\ref{E vac 2 electron-electron b massive}) and (\ref{E vac 2 electron-electron mix massive}) between two charged leptons with massive neutrinos, 
\begin{equation}
V_{\alpha \beta}(r) =  \frac{G_{F}^{2}}{4\pi^{3}r^{2}}
\sum_{a = 1}^{3} \left[m_{a}^{3} \left(g_{V}^{\alpha} +|U_{\alpha a}|^{2} \right)\left(g_{V}^{\beta} +|U_{\beta a}|^{2} \right) K_{3}(2m_{a}r) \right]
+ V_{\rm {\alpha \beta, mix}}(r),
\label{E vac 2 lepton lepton massive}
\end{equation}
where
\begin{equation}
\begin{split}
V_{\rm {\alpha \beta, mix}}(r) =    \frac{G_{F}^{2}}{4\pi^{3}r^{2}}
&  \sum_{a > b}^{3}  \frac{ \Re( U_{\alpha a}^* U_{\alpha b}^{}  U_{\beta b}^* U_{\beta a}^{} )  }{4}
\left\{
m_+^{ab} \left[\left(m_+^{ab}\right)^2 + \left(m_-^{ab}\right)^2 \right] K_3\left(\left.m_+^{ab}\right. r\right)
- \frac{4 \left(m_-^{ab}\right)^2 }{r} K_2\left(m_+^{ab}\, r\right)
\right. \\
 &
 \left. \mbox{}
+ \mathcal{O} \left[ \left(\frac{m_-^{ab}}{m_+^{ab}} \right)^2 \right] 
\right\}.
\end{split}
\label{E vac 2 lepton lepton mixing massive}
\end{equation}
	
Similarly, if the lightest neutrino is massless, the mixing term analogous to Eq.~(\ref{E vac 2 electron-electron b massless}) is given by

\begin{equation}
\begin{split}
V_{\rm {\alpha \beta, mix}}^{m_a=0}(r) = &   \sum_{\substack{b =1 \\ b \neq a}}^{3}
 \frac{ G_{F}^2 \Re( U_{\alpha a}^* U_{\alpha b}^{}  U_{\beta b}^* U_{\beta a}^{} )  } {48 \pi^3 r^5} \left[ 
e^{-m_b r} \left( 24 + 24 m_b r + 6 m_b^2 r^2 -2  m_b^3 r^3 +  m_b^4 r^4 - m_b^5 r^5 \right) \right.\\
& \left. -  (6 m_b^4 r^4 + m_b^6 r^6) \ \mathrm{Ei}\left(-m_b r\right) - 6 m_b^4 r^4 \ \Gamma\left( 0, m_b r \right) \right].
\end{split}   
\label{E vac 2 lepton lepton massless}
\end{equation}

\begin{figure}[t]
	\centering
	\includegraphics[width=10cm]{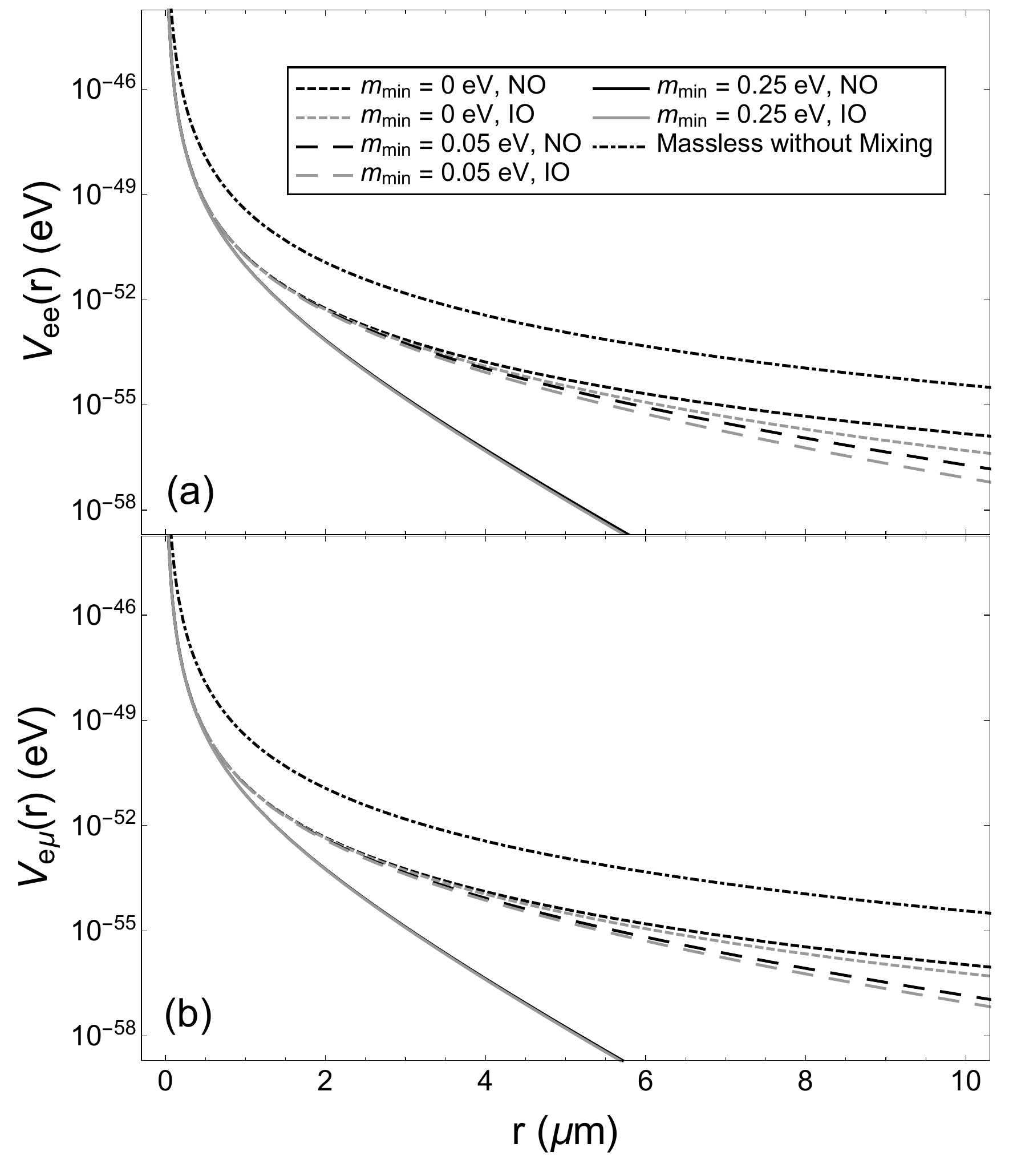}	
	\caption{Plots of the total 2NEP between (a) 2 electrons  and (b)  electron and muon.  Black lines represent normal ordering (NO) of neutrino masses while the gray lines represent inverted ordering (IO).  It is assumed respectively that $m_{\rm min}=0 \ \textrm{eV}$ (short-dashed), $m_{\rm min}=0.05\ \textrm{eV}$ (long-dashed), $m_{\rm min}=0.25\ \textrm{eV}$ (solid),  and that neutrinos are massless without mixing (black, dot-dashed). Values for all other parameters were obtained from the Particle Data Group \cite{PDG 2018} given in Table~\ref{PMNS table}.}
	\label{fig potential plot}
\end{figure}

\section{Discussion of Lepton-Lepton Results}

\subsection{Lepton-Lepton 2NEPs}

While the mixing of neutrino mass states does modify the 2NEPs involving leptons and nucleons, the most important effects are seen in interactions involving two leptons.   The consequences of the 2NEPs involving two leptons derived in the previous section  are explored  in Figs.~\ref{fig potential plot}, \ref{fig ratio mixing}, and \ref{fig ratio CP} using current neutrino parameter values from the Particle Data Group (Table~\ref{PMNS table}).   We consider normal ordering  (NO) of neutrino mass states ($m_{1} < m_{2} < m_{3}$) and inverted ordering (IO) ($m_{3} < m_{1} < m_{2}$).  Overall results for the lepton-lepton 2NEP with mixing obtained from Eqs.~(\ref{E vac 2 electron-electron b massive}), (\ref{E vac 2 electron-electron b massless}), (\ref{E vac 2 lepton lepton massive}) and (\ref{E vac 2 lepton lepton massless}) are plotted in Fig.~\ref{fig potential plot} for three different values of the lightest neutrino mass with both NO and IO: $m_{\rm min} = 0~{\rm eV}, 0.05~{\rm eV}$, and 0.25~eV. [The cases where the minimum neutrino mass state  $m_{\rm min} = 0 \ \textrm{eV}$ (short-dashed lines) with NO (black lines) and IO (gray lines) are usually referred to in the literature \cite{bilenky mass hierarchy} as the normal and inverted mass hierarchy, respectively, while the cases where $m_{\rm min} = 0.25 \ \textrm{eV}$ (black and gray solid lines) are examples of the quasi-degenerate scenarios.]  We see that the general behavior of the 2NEP with neutrino mixing does not differ significantly from the case without mixing. In particular, we see that they remain purely repulsive over all distances and fall-off drastically at large distances with the heaviest neutrino mass state determining the effective range of the interaction. At short separations ($r \ll 1/m_{a}$ for all $a$), the 2NEP behaves as if neutrinos are massless without mixing.  The difference between NO and IO  increases with  separation, but as the mass of the lightest neutrino increases, this difference quickly vanishes. It is also interesting to note that across all  masses of the lightest neutrino and all distances, IO produces smaller 2NEPs than with NO.

\subsection{Mixing between Different Neutrinos}
  
\begin{figure}[tb]
\centering
\includegraphics[width=15.5cm]{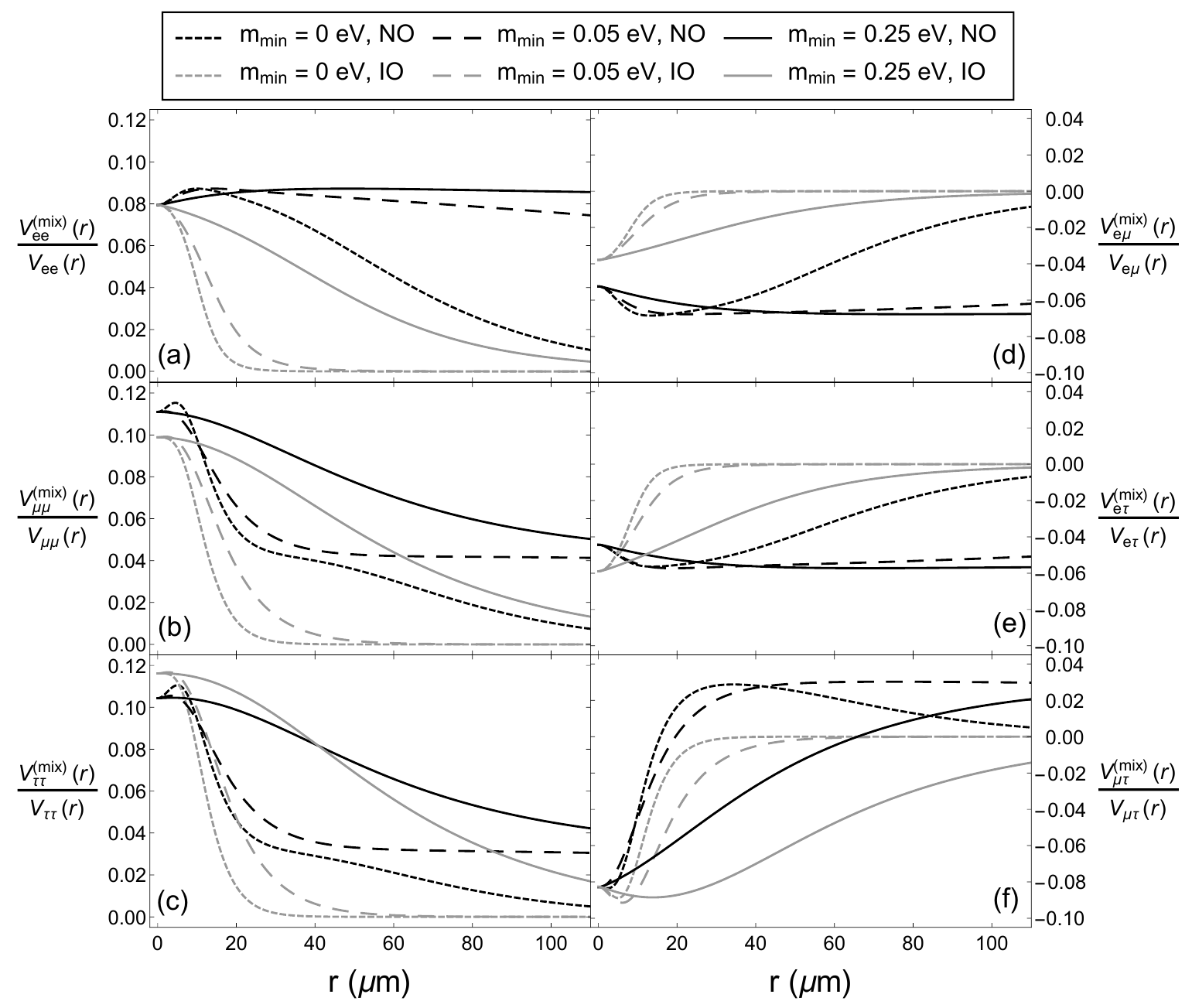}	
\caption{The ratios of the contribution from neutrino mixing to the overall 2NEP between: (a) Two electrons, (b) two muons, (c) two taus, (d)  electron and muon, (e) electron and tau and (f) muon and tau. Black lines represent normal ordering (NO) of neutrino masses while the gray lines represent inverted ordering (IO). It is assumed respectively that $m_{\rm min}=0 \ \textrm{eV}$ (short-dashed), $m_{\rm min}=0.05\ \textrm{eV}$ (long-dashed), $m_{\rm min}=0.25\ \textrm{eV}$ (sol id).  Values for all other relevant parameters were obtained from the Particle Data Group \cite{PDG 2018} given in Table~\ref{PMNS table}.}
\label{fig ratio mixing}
\end{figure}

The relative ratios of the mixing portion $V_{\alpha\beta}^{(\textrm{mix})}(r)$  of the 2NEP, arising from exchanging different neutrinos, to the total 2NEP between leptons $V_{\alpha\beta}(r)$ exhibited in Fig.~\ref{fig ratio mixing} shows a rich and interesting variety of behavior.  Generally, the reason is that  Eqs.~(\ref{E vac 2 lepton lepton massive}) and (\ref{E vac 2 lepton lepton massless}) involve sums over different  decaying terms with different characteristic length scales, coming from the sum of any two neutrino mass states, and the various combinations of mixing matrix parameters and the coupling constants.  Using Standard Model  parameters from the Particle Data Group \cite{PDG 2018}, we show that the new additional effect from neutrino mixing can get as large as nearly $12 \%$ of the overall 2NEP. In general, the relative strength of the mixing portion falls off at large distances, but there is no universal cut-off length scale to characterize this damping behavior. Most prominently, the effect of mixing seems to persist at much larger distances in NO than in IO. Within the Standard Model, due to the unitarity of the mixing matrix, one can see from Eqs.~(\ref{E vac 2 lepton lepton massive}) and (\ref{E vac 2 lepton lepton massless}) that the contribution from neutrino mixing is always positive for two leptons in the same generation, as seen in Fig.~\ref{fig ratio mixing}(a)--(c)  However, the mixing contribution in the 2NEP between leptons in different generations,  Fig.~\ref{fig ratio mixing}(d)--(f), shows a much richer behavior. In contrast to the 2NEP between leptons in the same generation, the mixing portion can alternate between positive and negative contribution at different length scales, and when the exchange of different mass states is involved, the mixing contribution can be negative for a wide range of separations.

\subsection{Effect of the Dirac \boldmath $CP$-Violating Phase}

\begin{figure}[t]	
\centering
\includegraphics[width=10cm]{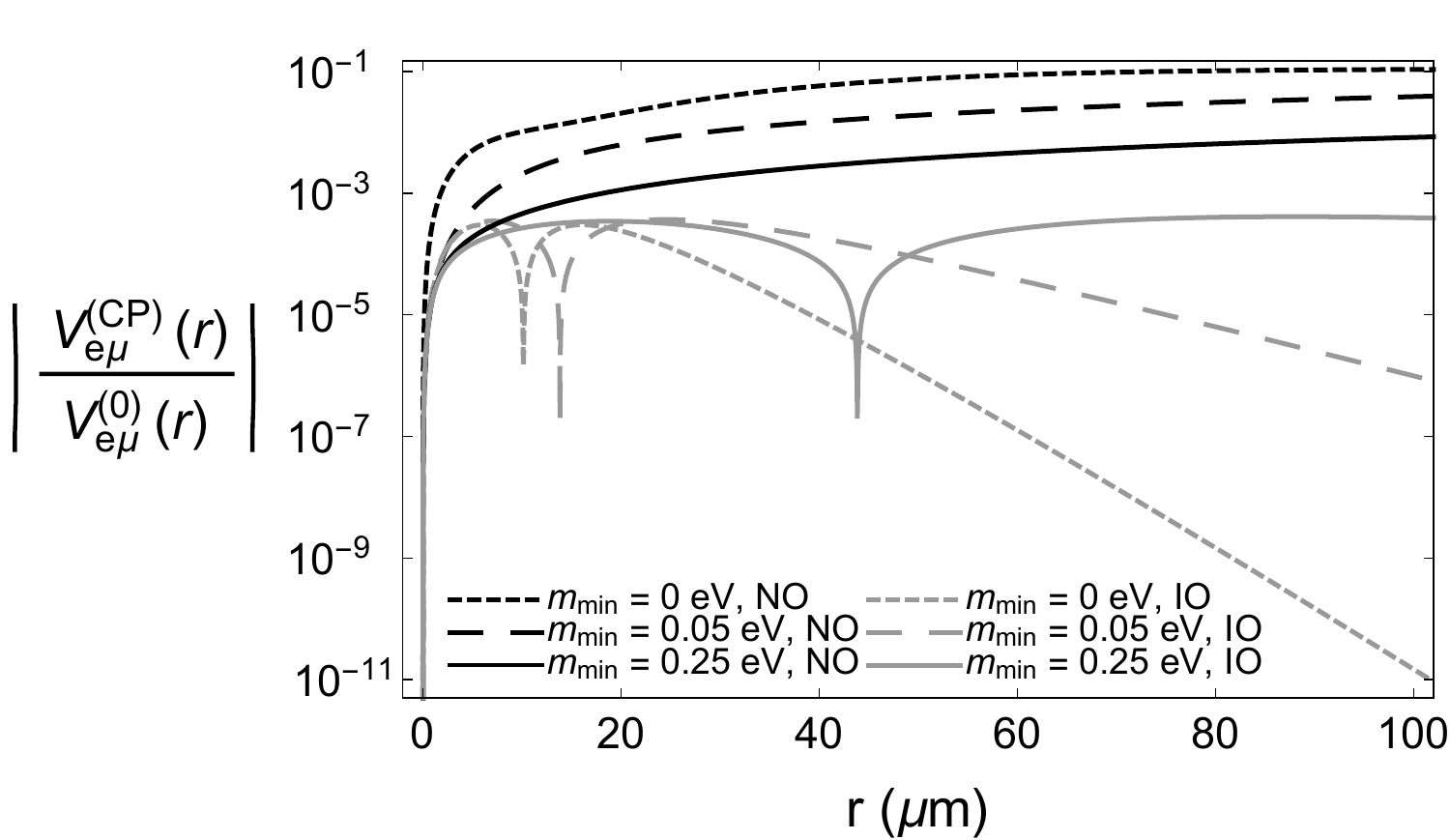}
\caption{The absolute value of the ratio  $V_{e\mu}^{(CP)}(r)/V_{e\mu}^{(0)}(r)$ from Eq.~(61) for several possibilities of minimum neutrino masses.  Black lines represent normal ordering (NO) of neutrino masses while the gray lines represent inverted ordering (IO).  It is assumed respectively that $m_{\rm min}=0 \ \textrm{eV}$ (short-dashed), $m_{\rm min}=0.05\ \textrm{eV}$ (long-dashed), $m_{\rm min}=0.25\ \textrm{eV}$ (solid). Values of all other  parameters were obtained from the Particle Data Group \cite{PDG 2018} given in Table~\ref{PMNS table}. Note: For $r>0$, each cusp in the IO curves corresponds to where the contribution of the $CP$-violating term changes its sign. }
\label{fig ratio CP}
\end{figure}

There is growing evidence that $CP$ is violated in neutrino oscillations which is reflected in a nonzero phase $\delta_{CP}$ in the PNMS matrix \cite{Abe CP}.    If this result is confirmed, the $CP$-violation will contribute to the 2NEP involving leptons, except the electron-electron 2NEP.  In the latter case, the matrix element $U_{\rm ea}$ involves an overall phase, which is not the case for the other lepton interactions.  To isolate the effects of the $CP$-violating phase $\delta_{CP}$, one can rewrite Eqs.~(\ref{E vac 2 lepton lepton massive}) and (\ref{E vac 2 lepton lepton massless}) into the simple form
\begin{equation}
V_{\alpha\beta}(r) = V_{\alpha\beta}^{(0)}(r) + V_{\alpha\beta}^{(CP)}(r)\, \sin^{2}\left(\frac{ \delta_{CP}}{2}\right),
\label{V CP}
\end{equation} 
where  $V_{\alpha\beta}^{(0)}(r)$ and $V_{\alpha\beta}^{(CP)}(r)$ are complicated functions that are independent of $\delta_{CP}$ and $V_{\rm ee}^{(CP)}(r) = 0$.  As an example, the ratios of $|V_{e\mu}^{(CP)}(r)/V_{e\mu}^{(0)}(r)|$ for various values of the lightest neutrino mass are plotted in Fig.~\ref{fig ratio CP}. For both NO and IO, the $V_{\alpha\beta}^{(CP)}(r)$  vanishes at short-range, but their behaviors at long-range are completely opposite.  While the contribution of the $CP$-violating term with NO is always positive, increases monotonically with distance and can get as large as $10 \%$ of the 2NEP, for IO, the $CP$-violating term changes its sign and falls off quickly at large distances.

\section{Summary and Future Directions}

To summarize, we have developed an alternative approach to derive the two-neutrino exchange potential (2NEP) between two fermions based on expressing the neutrino fields in terms of the Schr\"{o}dinger picture and finding the change in vacuum energy using time-independent perturbation theory.  We incorporated mixing of three neutrino flavors using the PMNS matrix $U_{\alpha a}$, where $\alpha$ denotes the neutrino flavor and $a$ denotes the neutrino mass state, and assumed the vacuum state was the tensor product of the individual mass vacuum states.  We were able to find  analytical expressions for the 2NEP for the nucleon-nucleon potential [Eq.~(\ref{general nucleon V})], the nucleon-lepton potential [Eq.~(\ref{general nucleon lepton V}], and the lepton-lepton potential when the lightest neutrino is massless [Eq.~(\ref{E vac 2 lepton lepton massless})].  While we were unable to find an analytical expression for the general lepton-lepton 2NEP, we obtained an exact result that can be evaluated numerically, and an approximate analytic formula which is quite precise for the range of neutrino masses consistent with current observations.  We were also able to isolate and evaluate the contribution of the $CP$-violating phase to the lepton-lepton 2NEP.  Where comparison is possible, our results agree with previous work.

While there was recent work which suggested the observation of the effects of the 2NEP might be within reach of spectroscopy experiments \cite{Stadnik}, followup calculations indicate this is unlikely \cite{Asaka}.  Fischbach, et al., showed that the 2NEP contribution to nuclear binding energy is of interest to precise tests of the weak interaction with respect to the equivalence principle  \cite{Fischbach PRD} and to lower limits on the neutrino masses from neutron star self-energies  \cite{Fischbach AoP}, these problems involve the nucleon-nucleon 2NEP which is unaffected by neutrino mixing within the Standard Model.  Experiments involving two leptons could observe the most interesting aspects of mixing with the 2NEP, but the observation of $CP$ violation in the 2NEP would require using  lepton-lepton systems other those involving only electrons (e.g., muonium).

In this paper we have focused our attention on the spin-independent 2NEP, but  the work by Stadnik \cite{Stadnik} highlights the importance of the spin-dependent 2NEP in realistic problems.  While we have assumed the Standard Model in our work, the mixing of neutrino mass states has raised the possibility of alternative vacuum states which would likely modify the 2NEP \cite{BV AoP,BV PLB}.  Recently, Blasone, et al., have studied the Casimir force between two plates assuming mixing of scalar fields for different vacua \cite{Blasone Casimir}.
We also assumed the neutrinos were Dirac neutrinos rather than Majorana neutrinos.  In addition, recent experiments and cosmological observations hint at the possibility of sterile neutrinos which would also impact the 2NEP \cite{Sterile neutrinos}.  The 2NEP is interesting because it probes fundamental issues of neutrino physics such as the neutrino mass and mixing, the number of neutrinos, the type of neutrino (Dirac or Majorana), $CP$-violation, the neutrino vacuum state, while producing a result, an interaction potential, that is familiar to an introductory physics student.  One can only hope that someday direct evidence of the 2NEP will be observed in nature.

\appendix

\section{Calculation of Single Flavor two-neutrino Potential}
\label{Calculation Appendix}

In this appendix, we calculate the second-order energy shift of the single neutrino vacuum by two fermions which depends on the  fermion separation, Eq.~(\ref{E vac 2c}),
\begin{equation}
E_{\rm vac}^{(2)}(\vec{r}_{1}-\vec{r}_{2}) = 
	-\sum_{\vec{k}',\vec{k}} \sum_{s,s'}\left\{
	\frac{\left[\langle 0|H_{{\rm int},1}|\vec{k}',s'\rangle_{\nu}|\vec{k},s\rangle_{\bar{\nu}}\right] 
	\left[_{\bar{\nu}}\langle \vec{k},s| \,_{\nu}\langle \vec{k}',s'|H_{{\rm int},2}|0\rangle\right]}{\omega_{\vec{k'}} + \omega_{\vec{k}}} + {\rm c.c.}\right\}.
\label{E vac 2c A}
\end{equation}
Using Eq.~(\ref{H int i}), the required matrix element is
\begin{equation}
_{\bar{\nu}}\langle \vec{k},s| \,_{\nu}\langle \vec{k}',s'|H_{{\rm int},i}|0\rangle =
	  \frac{G_{F}g_{V,i}^{f}}{\sqrt{2}} \frac{m_{\nu}}{V}\frac{1}{\sqrt{\omega_{\vec{k}}\omega_{\vec{k}'}}}
	\,u^{\dag}_{s'}(\vec{k}')\left(1 - \gamma^{5}\right) v_{s}(\vec{k})e^{-i(\vec{k}' + \vec{k}) \cdot \vec{r}_{i}}.
\label{H int i matrix}
\end{equation}
Substituting Eqs~(\ref{H int i matrix}) and its complex conjugate into Eq.~(\ref{E vac 2c}) then gives
\begin{equation}
E_{\rm vac}^{(2)}(\vec{r}_{1}-\vec{r}_{2}) = 
	-g_{V,1}^{f}g_{V,2}^{f}\left( \frac{G_{F}}{\sqrt{2}}\frac{m_{\nu}}{V}\right)^{2}\sum_{\vec{k}',\vec{k}} \sum_{s,s'}
	\left\{
	\frac{\left[v^{\dag}_{s'}(\vec{k})\left(1 - \gamma^{5}\right)u_{s'}(\vec{k}')\right]\left[u^{\dag}_{s'}(\vec{k}') \left(1 - \gamma^{5}\right)v_{s}(\vec{k})\right]e^{-i(\vec{k}' + \vec{k}) \cdot (\vec{r}_{1} - \vec{r}_{2})}}{\omega_{\vec{k}}\omega_{\vec{k}'}\left(\omega_{\vec{k'}} + \omega_{\vec{k}}\right)} + {\rm c.c.}\right\}.
\label{E vac 2d}
\end{equation}
Using
\begin{equation}
\sum_{s,s'}\left[v^{\dag}_{s'}(\vec{k})\left(1 - \gamma^{5}\right)u_{s'}(\vec{k}')\right]\left[u^{\dag}_{s'}(\vec{k}')\left(1 - \gamma^{5}\right)v_{s}(\vec{k})\right] =	\frac{2}{m_{\nu}^{2}}\left( \omega_{\vec{k}}\omega_{\vec{k}'} + \vec{k}\cdot\vec{k'} \right),
\label{vuuv}
\end{equation}
one can show that Eq.~(\ref{E vac 2d}) can be written as
\begin{equation}
E_{\rm vac}^{(2)}(\vec{r}_{1}-\vec{r}_{2}) = -2g_{V,1}^{f}g_{V,2}^{f}G_{F}^{2}\left(\frac{1}{V}\right)^{2}
	\sum_{\vec{k}',\vec{k}}
	\left[\frac{\omega_{\vec{k}}\omega_{\vec{k}'} + \vec{k}\cdot \vec{k}'}{\omega_{\vec{k}}\omega_{\vec{k}'}\left(\omega_{\vec{k'}} + \omega_{\vec{k}}\right)}\right] 
	e^{i(\vec{k}' + \vec{k}) \cdot (\vec{r}_{1} - \vec{r}_{2})}.
\label{E vac 3e}
\end{equation}

To evaluate the sums in Eq.~(\ref{E vac 3e}), we go to the continuum limit, which gives
\begin{equation}
E_{\rm vac}^{(2)}(\vec{r}) = -\frac{2\,g_{V,1}^{f}g_{V,2}^{f}G_{F}^{2}}{(2\pi)^{6}}\int d^{3}k\, \int d^{3}k' 
\left\{
\left[\frac{\omega_{\vec{k}}\omega_{\vec{k}'} + \vec{k}\cdot \vec{k}'}{\omega_{\vec{k}}\omega_{\vec{k}'}\left(\omega_{\vec{k'}} + \omega_{\vec{k}}\right)}\right] e^{i(\vec{k}' + \vec{k}) \cdot \vec{r}}\right\},
\label{E vac 2f}
\end{equation}
where  $\vec{r} \equiv \vec{r}_{1} - \vec{r}_{2}$.    Rather than directly evaluating the integrand in Eq.~(\ref{E vac 2f}), we will first replace the term  involving $\omega_{\vec{k}}\omega_{\vec{k}'} $ in the numerator with
\begin{equation}
\omega_{\vec{k}}\omega_{\vec{k}'} \, e^{i(\vec{k}' + \vec{k}) \cdot \vec{r}}
	= 
\left[\frac{1}{2}\left(\omega_{\vec{k}} + \omega_{\vec{k}'}\right)^{2} - m_{\nu}^{2} + \vec{k}\cdot \vec{k}' 
		+ \frac{1}{2}\vec{\nabla}^{2} \right] e^{i(\vec{k}' + \vec{k}) \cdot \vec{r}}.
\label{integral evaluation trick}
\end{equation}
Then
\begin{equation}
E_{\rm vac}^{(2)}(\vec{r}) = -\frac{2\,g_{V,1}^{f}g_{V,2}^{f}G_{F}^{2}}{(2\pi)^{6}}\int d^{3}k\, \int d^{3}k' 
\left\{
\left[\frac{\frac{1}{2}\left(\omega_{\vec{k}} + \omega_{\vec{k}'}\right)^{2} -  m_{\nu}^{2} + 2\vec{k} \cdot \vec{k}'
		+ \frac{1}{2}\vec{\nabla}^{2}}{\omega_{\vec{k}}\omega_{\vec{k}'}\left(\omega_{\vec{k'}} + \omega_{\vec{k}}\right)}\right] e^{i(\vec{k}' + \vec{k}) \cdot \vec{r}}\right\},
\label{E vac 2g}
\end{equation}
which can be rewritten in terms of four separate integrals,
\begin{equation}
E_{\rm vac}^{(2)}(\vec{r}) = -g_{V,1}^{f}g_{V,2}^{f}G_{F}^{2}\left[I_{1}(\vec{r}) + I_{2}(\vec{r}) + 4I_{3}(\vec{r}) + I_{4}(\vec{r})\right],
\label{E vac 2h}
\end{equation}
given by
\begin{eqnarray}
I_{1}(\vec{r}) & = & -\frac{2m_{\nu}^{2}}{(2\pi)^{6}}\int d^{3}k\, \int d^{3}k' \left[\frac{e^{i(\vec{k}' + \vec{k}) \cdot \vec{r}}}{\omega_{\vec{k}}\omega_{\vec{k}'}\left(\omega_{\vec{k'}} + \omega_{\vec{k}}\right)}\right]  =  -\frac{2m_{\nu}^{3}}{8\pi^3 r^{2}} K_{1}(2m_{\nu}r),
\label{I1} \\
I_{2}(\vec{r}) & = & \vec{\nabla}^{2}\left\{
	\frac{1}{(2\pi)^{6}}\int d^{3}k\, \int d^{3}k' \left[\frac{e^{i(\vec{k}' + \vec{k}) \cdot \vec{r}}}{\omega_{\vec{k}}\omega_{\vec{k}'}\left(\omega_{\vec{k'}} + \omega_{\vec{k}}\right)}\right] \right\} = 
	 \frac{2m_{\nu}}{8\pi^{3}r^{4}} \, \left[3m_{\nu}rK_{0}(2m_{\nu}r) + (3 + 2 m_{\nu}^{2}r^{2})K_{1}(2m_{\nu}r)\right],
\label{I2} \\
I_{3}(\vec{r}) & = & \frac{1}{(2\pi)^{6}}\int d^{3}k\, \int d^{3}k' \left[\frac{(\vec{k} \cdot \vec{k}') \,e^{i(\vec{k}' + \vec{k}) \cdot \vec{r}}}{\omega_{\vec{k}}\omega_{\vec{k}'}\left(\omega_{\vec{k'}} + \omega_{\vec{k}}\right)}\right]  \nonumber \\
& = & -\frac{m_{\nu}}{32\pi^{3}r^{4}}\left[4m_{\nu}r K_{0}(2m_{\nu}r) + (4 + 3m_{\nu}^{2}r^{2})K_{1}(2m_{\nu}r) 
	+ 4m_{\nu}rK_{2}(2m_{\nu}r) + m_{\nu}^{2}r^{2}K_{3}(2m_{\nu}r) \right],
\label{I3} \\
I_{4}(\vec{r}) & = & \frac{1}{(2\pi)^{6}}\int d^{3}k\, \int d^{3}k' \left[\frac{\left(\omega_{\vec{k}} + \omega_{\vec{k}'}\right) e^{i(\vec{k}' + \vec{k}) \cdot \vec{r}}}{\omega_{\vec{k}}\omega_{\vec{k}'}}\right] =  \frac{m_{\nu}}{\pi^{2}r} K_{1}(m_{\nu}r)\,  \delta^{3}(\vec{r}).
\label{I4}
\end{eqnarray}
All four of these integrals  depend only on the particle separation $r$ which is required by spatial isotropy and translation invariance.  Since we assume $r > 0$, the divergent contact contribution arising from $I_{4}(\vec{r})$ will be dropped.  (Our low-energy theory certainly breaks down at small separations as discussed earlier.)   Combining Eqs.~(\ref{E vac 2h})--(\ref{I3}) gives our final result for the two-neutrino exchange potential,
\begin{equation}
V_{\nu,\bar{\nu}}(r) = \frac{G_{F}^{2}g_{V,1}^{f}g_{V,2}^{f}m_{\nu}^{3}}{4\pi^{3}r^{2}}K_{3}(2m_{\nu}r).
\end{equation}

\acknowledgments

We thank Ephraim Fischbach for insightful discussions and for  his earlier work which motivated this paper, and the anonymous referee for comments which greatly improved our presentation.  We also thank Wabash College for providing summer stipends which supported some of this work.

\end{document}